\documentclass[reprint,pra,aps,twocolumn,superscriptaddress]{revtex4-2}
\usepackage{bm}
\usepackage{graphicx}
\usepackage{times}
\usepackage{soul}
\usepackage{color}
\usepackage[colorlinks, citecolor = blue, linkcolor = blue, urlcolor  = blue, anchorcolor = blue, breaklinks = true]{hyperref}
\usepackage{mathtools}
\usepackage{array}
\usepackage{amsmath}
\usepackage{bbold}

\def \FUW{Institute of Experimental Physics, Faculty of Physics, University of Warsaw, Pasteura 5, 02-093 Warsaw, Poland}

\begin{document}

\title{Single vanadium ion magnetic dopant in an individual CdTe/ZnTe quantum dot}

\author{K. E. \surname{Po\l{}czy\'nska}}\affiliation{\FUW}\email{kpolczynska@fuw.edu.pl, k.e.polczynska@gmail.com}
\author{T. \surname{Kazimierczuk}}\affiliation{\FUW}
\author{P. \surname{Kossacki}}\affiliation{\FUW}
\author{W. \surname{Pacuski}}\affiliation{\FUW}

\date{\today}

\begin{abstract}
We present the basic properties of a new physical system: an individual V$^{2+}$ ion embedded into an individual quantum dot. The system is realized utilizing molecular beam epitaxy and it is observed using a low-temperature polarization-resolved magneto-photoluminescence. The nature of the system is confirmed by observation of the excitonic lines split due to the interactions of a vanadium ion with carriers confined in a CdTe/ZnTe quantum dot. Observed data are explained by the numerical modeling which includes $s{,}p$-$d$ exchange interaction, Zeeman splitting of the exciton and the ion, diamagnetic shift, and the presence of shear strain within the quantum dot. The fundamental state of vanadium exhibits a spin $\pm \frac{1}{2}$ making this system a textbook localized qubit.

\end{abstract}

\maketitle

\section{Introduction}

An individual epitaxial quantum dot (QD) with narrow excitonic lines serves as an excellent tool for studying embedded magnetic dopants and their interactions with excitons via the $s{,}p$-$d$ exchange. This $s{,}p$-$d$ exchange interaction causes excitonic line splitting into multiple components, determined by the magnetic ion’s spin and various interactions within the QD \cite{besombes_2004_prl}. The first observed individual magnetic ion in a QD was manganese \cite{besombes_2004_prl,Kudelski_2007_PRL,LeGall_2009_PRL,Goryca_2009_PRL2,Goryca_PRL_2014,Lafuente_Sampietro_2015_PRB,Fainblat_NL_2016,Kasprzak2022,PhysRevB.107.235305}, which exhibits a spin of $\frac{5}{2}$, the maximal value for a $d$-shell. Subsequent studies reported ions with spins of $\frac{3}{2}$ (cobalt \cite{kobak_2014_natcom,Kobak2018,PhysRevB.109.235302}) and 2 (iron \cite{smolen_fe_2016_natcom,Smolenski_2017} and chromium \cite{Lafuente_2016,Lafuente_APL_2016,besombes_2019_prb,Tiwari2020,Tiwari2021,Tiwari2022,PhysRevB.106.045308}). However, the simplest case of a magnetic ion with a spin of $\frac{1}{2}$ has long remained elusive. In this work, we demonstrate this case through the observation of an individual vanadium ion.

Previous research on vanadium dopants has primarily focused on their influence in bulk cadmium telluride \cite{wanad_cdte_christmann,wanad_cdte_peka,PhysRevB.53.3634,PhysRevB.49.5274}. In our study, we investigate self-assembled CdTe QDs doped with vanadium and embedded in a ZnTe barrier, fabricated using molecular beam epitaxy (MBE). We report the observation of a single quantum dot containing an individual vanadium dopant and measure its magneto-optical properties through micro-photoluminescence experiments. Numerical modeling of the experimental data reveals that the key phenomenon underpinning the observed spectral features is the presence of shear strain within the quantum dot. Ultimately, our results confirm that vanadium in this system exhibits a spin projection of $\pm \frac{1}{2}$, establishing our system as a realization of the canonical half-spin qubit.

\section{Technology}

\begin{figure*}
\includegraphics[width=\textwidth]{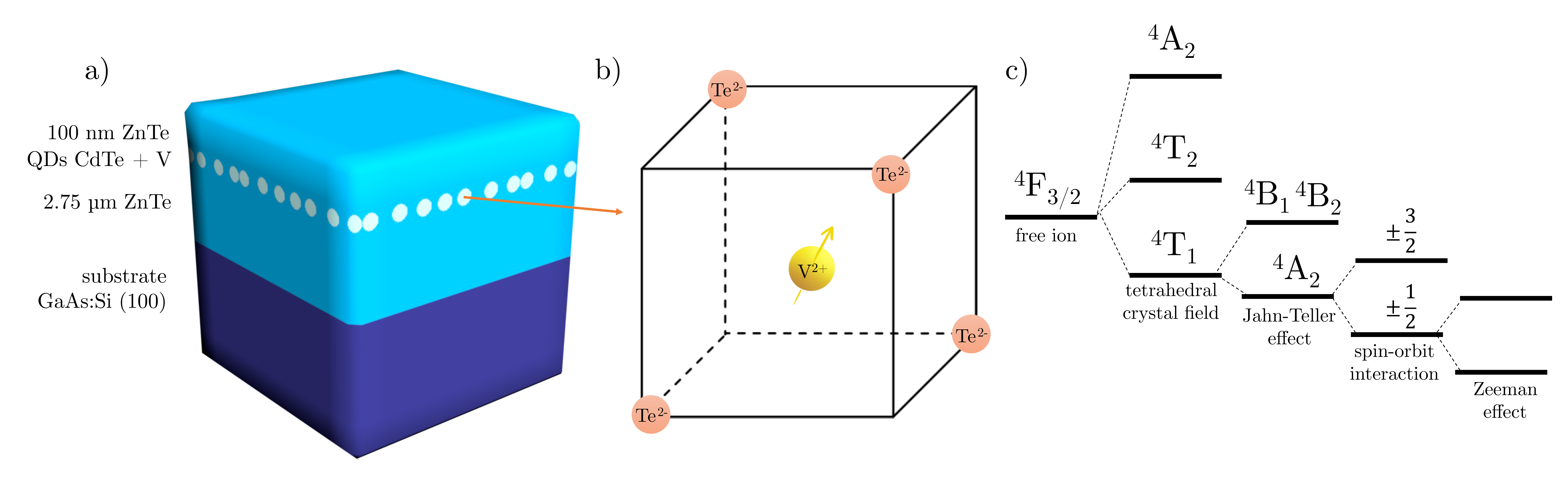}
\caption{\textbf{Vanadium as a dopant in the epitaxial QDs made of CdTe and ZnTe.} a) Scheme of the structure grown by MBE. b) Ion of vanadium in the tetrahedral symmetry of the closest neighbors of the tellurium anions. c) Scheme of the energy levels splitting by the influence of the crystal lattice, Jahn-Teller effect, spin-orbit coupling, and Zeeman effect. Atomic term symbols after \cite{wanad_cdte_christmann}.}
\label{scheme}
\end{figure*}

The schematic representation of the sample structures is shown in Fig. \ref{scheme}a. The samples were grown in an MBE chamber provided by SVT Associates. The growth process begins with a 3-inch diameter GaAs:Si (100) substrate with a $2^\circ$ offset. A 2 $\mu$m ZnTe buffer layer is deposited onto the substrate, and its surface is protected with amorphous tellurium before removal from the growth chamber. The substrate is then quartered to serve as the base for the growth of four samples.

The next step involves the deposition of a 730 nm ZnTe bottom barrier layer. Initially, the growth temperature is $370^\circ$C, then gradually suced to $350^\circ$C, where it is maintained for the remainder of the process. Following this, the shutters for the Cd, Te, and V sources are opened for 5 seconds, corresponding to an average deposited layer thickness of approximately 2 nm. However, as evidenced by the photoluminescence spectra provided in the Supplementary Information (SI), this deposition process leads to the formation of quantum dots (QDs). The deposition is followed by a 5-second exposure to a Te flux, after which a 100 nm ZnTe cap layer is grown.

This method results in a short quantum dot formation time, yielding a high density of uniform QDs \cite{Karczewski1999,Ragunathan2019}. We chose this approach to mitigate segregation effects, characteristic for more time-intensive methods, such as those where the CdTe layer is cooled and temporarily covered with amorphous Te \cite{Tinjod2003}. Such alternative methods tend to produce QDs with lower densities and broader size distributions \cite{Wojnar2007,wojnar-jcg-2011,Ragunathan2019}.

Using this methodology, a series of four samples is prepared, with vanadium concentration as the sole variable parameter. The dopant concentration is controlled by adjusting the temperature of the vanadium effusion cell, ranging from $1200^\circ$C to $1500^\circ$C in $100^\circ$C increments. Higher source temperatures result in increased vanadium concentrations within the quantum dots. Optical characterization of the samples reveals standard spectral properties of CdTe/ZnTe QDs \cite{Kobak_statistics_2013}, as presented in the Supplementary Materials. The data shows that intense and narrow QD emission lines can be observed only for the sample with the lowest vanadium concentration (see Fig. S1). For higher doping levels, photoluminescence from the nanostructures is suppressed, consistent with previous observations of transition-metal-doped structures, where efficient photoluminescence quenching dominates the optical properties \cite{Nawrocki1995,QuenchingVanadium,Oreszczuk2017,Piotrowski2020}. Consequently, the structure presented in this paper is grown with the vanadium source set to $1200^\circ$C.

\section{Experimental results}

Optical experiments were performed at a temperature of 1.6 K. The sample was mounted on the immersion microscope objective \cite{jasny1996} within a helium bath cryostat (Spectromag, Oxford Instruments) equipped with a superconducting coil capable of generating magnetic fields of up to 10 T. All measurements were conducted in the Faraday configuration (magnetic field along the optical axis, which was perpendicular to the sample surface). The QDs were excited using a 532 nm laser, and the excitation location was adjusted by shifting the laser spot through the movement of a lens positioned in front of the cryostat. Signal detection was polarization-resolved. A detailed schematic of the experimental setup is provided in Fig. S2.

The optical signal from the sample exhibits multiple narrow lines corresponding to different quantum dots (see Fig. S1). Single quantum dots are primarily identifiable in the low-energy region of the QD ensemble spectrum. Most of the observed single QDs display characteristic features typical of nonmagnetic, undoped CdTe/ZnTe QDs, as reported in previous studies \cite{Kobak_statistics_2013}. These features include a standard spectral pattern, with a consistent energy separation of approximately 10--15 meV between the neutral exciton (X) and the biexciton (XX). A positive trion (X$^+$) appears in the middle of this separation, and a negative trion (X$^-$) is observed between X$^+$ and XX.

Notably, the X and XX lines exhibit fine structure splitting (FSS) and are orthogonally linearly polarized with respect to each other. In contrast, the trion lines generally appear as single peaks without discernible linear polarization dependence (see Fig. S3). In the presence of a magnetic field, trions in undoped QDs exhibit Zeeman and diamagnetic shifts, while the X and XX lines display an additional feature. At zero magnetic field, X and XX formed doublets due to FSS, which split further into two circularly polarized components as the magnetic field increased. This behavior results in anticrossings at B = 0 T, as shown in Fig. S4.

To identify quantum dots containing a single vanadium ion, we initially search for isolated QDs at zero magnetic field, focusing on those with multiple excitonic spectral lines, line broadening, and abnormal linear polarization behavior. However, the definitive distinction between doped and undoped QDs is most visible in the presence of a magnetic field. In undoped QDs, only two anticrossings are observed, corresponding to X and XX at B = 0 T. In contrast, QDs containing a magnetic dopant exhibit a more complex spectrum characterized by multiple anticrossings.
\begin{figure*}
\includegraphics[scale=0.35]{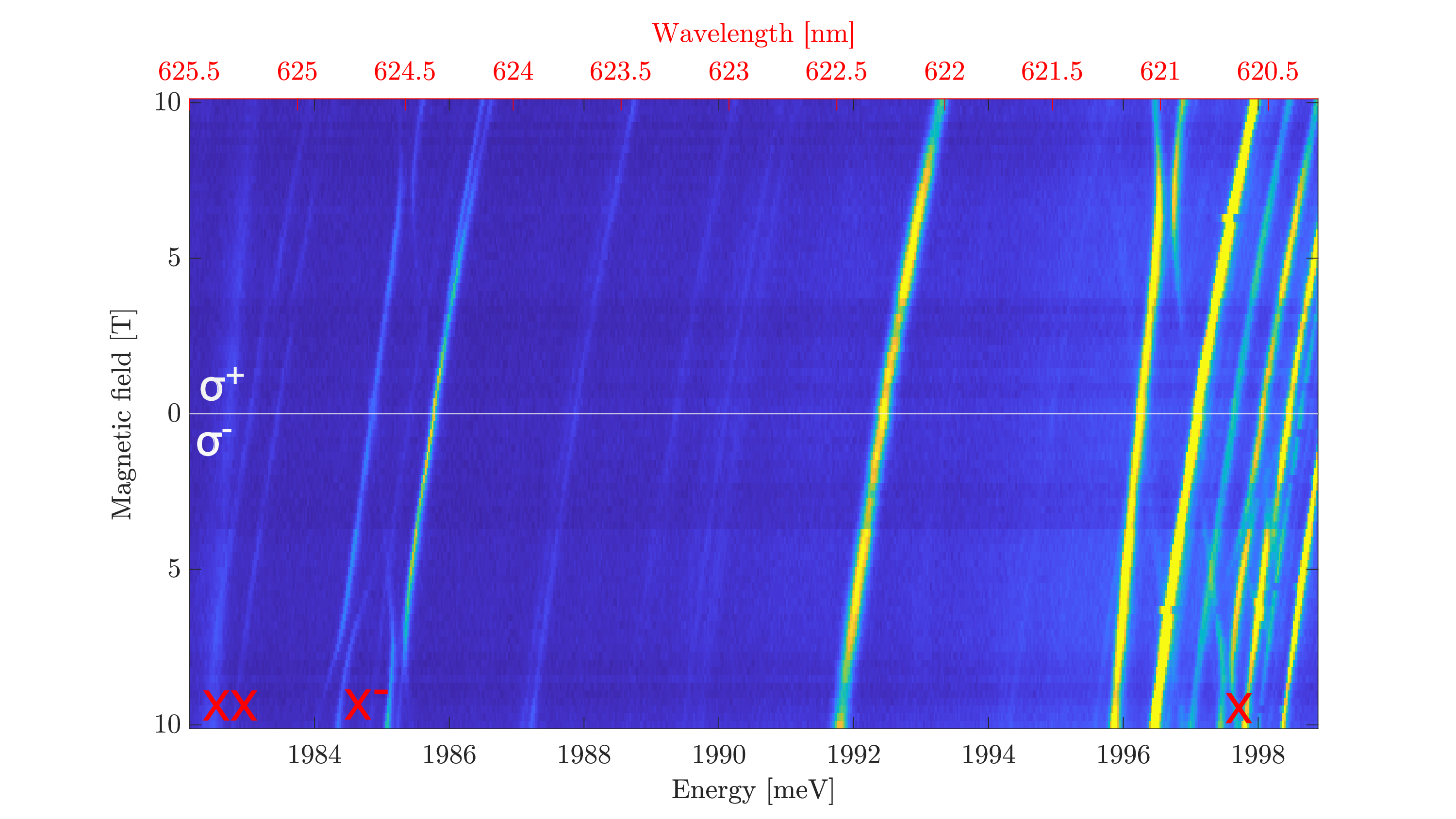}
\caption{\textbf{The first report of the QD with a solitary V$^{2+}$ dopant.} The measurement of polarization-resolved magnetospectroscopy revealed neutral exciton (X), biexciton (XX), and negative trion (X$^{-}$) of a CdTe QD consisting of an ion of vanadium. The experiment was performed at a temperature of $1.6$\,K.}
\label{measurement1}
\end{figure*}

\begin{figure*}
\includegraphics[width=\textwidth]{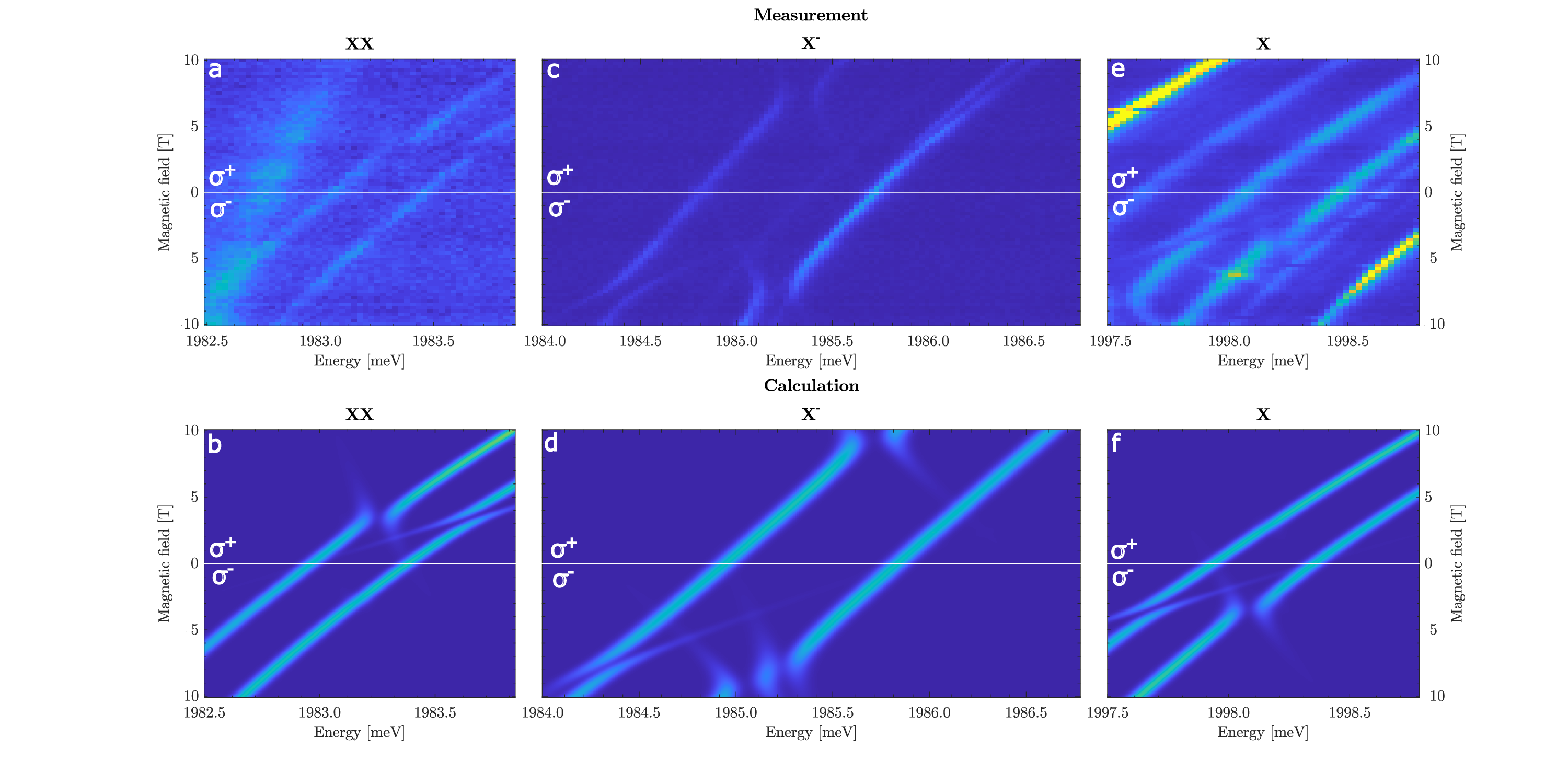}
\caption{\textbf{Excitonic complexes of a QD consisting V$^{2+}$: comparison of experimental (a, c, e) and calculated (b, d, f) results.} The spin Hamiltonian model included exange interaction between excitons and ion, Zeeman splitting of the exciton and the ion, diamagnetic shift, and the shear strain influence on the valence band. Final parameters of the Hamiltonian: $\Delta_{s{,}p-d} = 0.4$\,meV, $\delta_\mathrm{X} = 0.025$\,meV, $g_\mathrm{V} = 2$, $g_\mathrm{X} = 2.7$, $\gamma = 0.0015\,\frac{\mu \textnormal{eV}}{\textnormal{T}^2}$, $\Delta_{p-d}\xi = 0.0525$ meV, $T = 30$\,K, $k = 0.0862\,\frac{\textnormal{meV}}{\textnormal{K}}$, $\mu _b = 5.788\times10^{-2}\,\frac{\textnormal{meV}}{\textnormal{T}}$.}
\label{measurement}
\end{figure*}

Magnetospectroscopic measurements of a QD containing a single vanadium dopant are shown in Fig. \ref{measurement1}. Among the numerous lines observed, a subset originates from the same QD. All excitonic lines exhibit both Zeeman and diamagnetic shifts, with no significant angular dependence on the linear polarization of the photoluminescence (see Fig. S5). We identify three excitonic complexes: the neutral exciton, the negative trion, and the biexciton.

Each of these lines is split into two components in the absence of a magnetic field. The splitting values is the same for X and XX (0.4 meV) but larger for X$^-$($0.9$\,meV). Under an applied magnetic field, X and XX exhibit two anticrossings at a field of $3.75$\,T ($B_\mathrm{ac}$) in opposite circular polarisation ($\sigma ^-$ for X). For the X$^-$, three anticrossings are observed. A detailed explanation of these spectral features, supported by numerical calculations, will be provided in the next chapter.

Additionally, two more anticrossings are visible in Fig. \ref{measurement1}. One occurs in the lower-energy line of the neutral exciton (1998.8 meV in B = 0 T) at -9 T. The other is observed at 7.5 T in a single line located at 1996.23 meV in B = 0 T. These features are attributed to the dark exciton, which can become partially brightened through interactions with the magnetic ion and bright excitons \cite{besombes_2004_prl,Goryca_PRB_dark,Kobak2018}.

\section{Theoretical model}

The observed anticrossings and split lines in the examined quantum dot (QD) (Fig. \ref{measurement1}) can be attributed to the presence of a magnetic ion within the nanostructure. Vanadium in the +2 oxidation state (V$^{2+}$) substitutes for cadmium in the CdTe crystal lattice due to their identical number of valence electrons, thereby acting as an electrically neutral dopant, as illustrated in Fig. \ref{scheme}b. In this substitutional site, the vanadium ion is coordinated by four tellurium anions arranged in tetrahedral symmetry. This crystal field splits the energy levels into $^4A_2$, $^4T_2$, and the ground state $^4T_1$ \cite{wanad_cdte_christmann}, as depicted in Fig. \ref{scheme}c.

Considering the Jahn-Teller effect, the lowest energy level is $^4A_2$, which is further split by spin-orbit coupling into the $\pm \frac{3}{2}$ (higher energy) and $\pm \frac{1}{2}$ (lower energy) states. The subsequent model focuses solely on the ground state with a spin of $\pm \frac{1}{2}$. The calculations concentrate on the neutral exciton, representing the simplest case. This transition involves the QD state without charge carriers, thereby directly revealing the properties of a single vanadium ion.

In a system comprising an exciton and a magnetic atom within a QD, the typical phenomena considered include: $s{,}p$-$d$ interaction (1), anisotropy of the exciton (2), Zeeman splitting of both exciton (3) and ion (4), and diamagnetic shift (5) \cite{besombes_2004_prl,Kudelski_2007_PRL,kobak_2014_natcom, smolen_fe_2016_natcom, Lafuente_2016}. However, for a QD doped with a V$^{2+}$ ion, not all spectral features could be accounted for using this basic model. Due to the presence of significant valence band mixing, additional terms needed to be incorporated into the Hamiltonian. The most notable contribution to the system arise from the effects of shear strain (6) on the QD \cite{PhysRevB.104.L041301}. Only bright excitons were considered. The final Hamiltonian for the system is given as:

\begin{equation}
\nonumber
 \begin{multlined}
H_\mathrm{X+V} = - \left(\frac{1}{2}\Delta_{s{,}p-d} S_z \otimes \sigma_z \right)_1 + \left(\frac{1}{2}\delta_X \mathbb{1} \otimes \sigma_x\right)_2 + \\
- \left(\frac{1}{2} \mu_B g_X B\, \mathbb{1} \otimes \sigma_z\right)_3 - \left(\frac{1}{2} \mu_B g_V B S_z \otimes \mathbb{1}\right)_4 + \\
+ \left(\frac{1}{2} \gamma^2 B\, \mathbb{1} \otimes \mathbb{1} \right)_5 + H_{\mathrm{strain},6}
 \end{multlined}
 \label{hamiltonianV}
\end{equation}

where $\Delta_{s{,}p-d}$ is a constant describing the exciton-ion exchange interaction, $\delta_\mathrm{X}$ represents the splitting of energy levels due to QD anisotropy, and $\gamma$ is the excitonic diamagnetic shift constant. The symbols$\mu_B$, $g_\mathrm{X}$ and $g_\mathrm{V}$ denote the Bohr magneton and the g-factors of the exciton and vanadium ion, respectively. The spin operators for the ion ($S$) and exciton ($\sigma$) are treated in their standard forms. The term $H_\mathrm{strain}$ introduces additional contributions arising from the inferred shear strain. As detailed in the Supplementary Information of Ref. \cite{PhysRevB.104.L041301}, this form of strain facilitates spin flips of the vanadium ion, with a probability proportional to the z-component of the hole wavefunction in the excitonic state. The explicit matrix representation of this operator is provided in the Supplementary Information.

\begin{figure}[hbt!]
\includegraphics[width=0.5\textwidth]{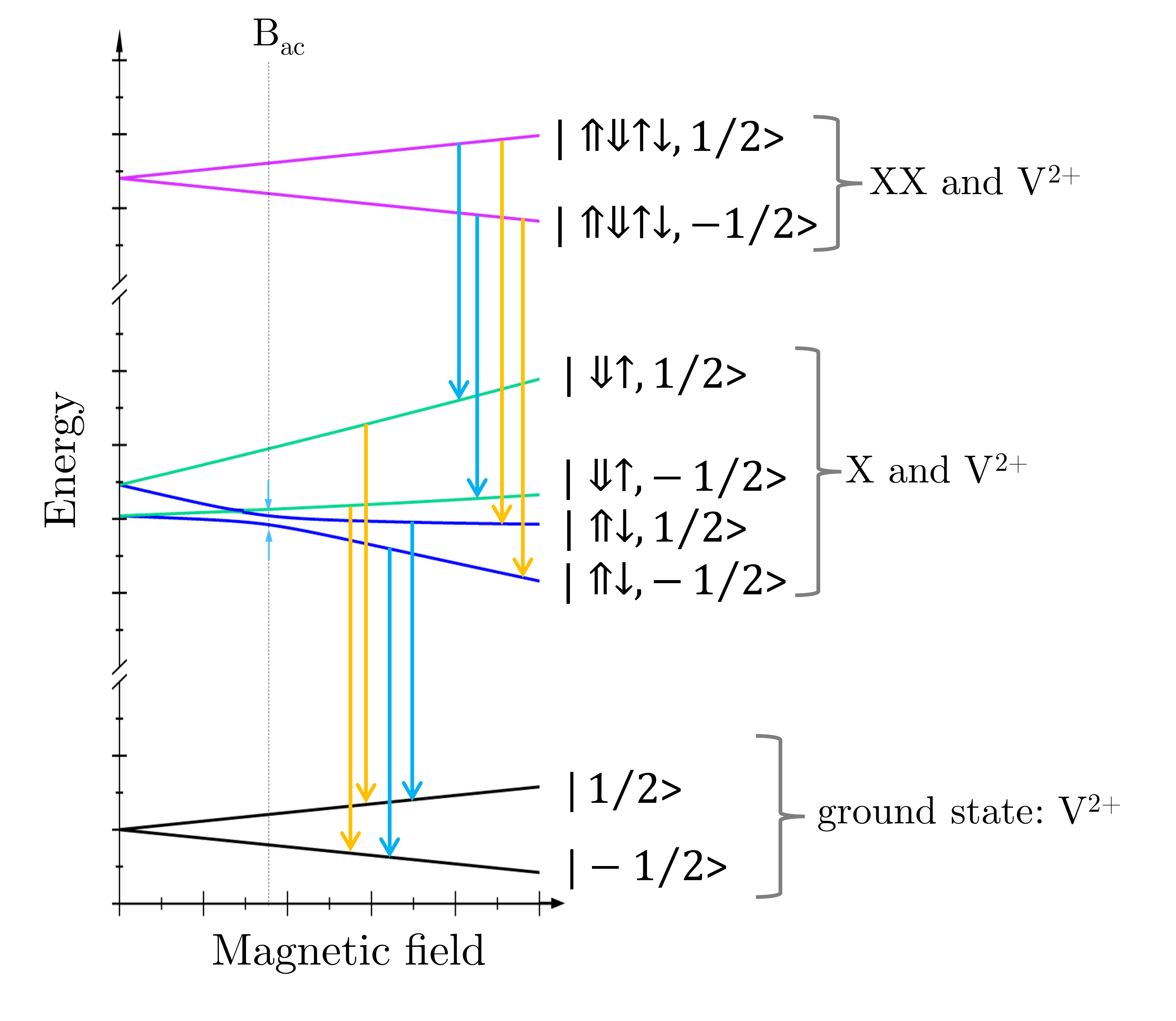}
\caption{\textbf{Energy level scheme of the XX-X cascade of the QD with single vanadium dopant.} The origin of the anticrossings observed in the magnetic field $B_\mathrm{ac}$ is located in the X, in the splitting of the state with the same orientation of the hole spin, but a different orientation of the spin of vanadium (dark blue lines). Optical transitions in $\sigma^-$ polarization are marked with light blue arrows, while orange arrows represent optical transitions observed in $\sigma^+$ polarization. }
\label{levels}
\end{figure}

Spin Hamiltonians for X$^{-}$ and XX in the QD containing a single vanadium dopant are calculated analogously to the neutral exciton case. The results of these calculations are compared with experimental data in Fig. \ref{measurement}. A key feature observed for all excitonic complexes is the presence of anticrossings, appearing at the same magnetic field values. For the neutral exciton, anticrossings and cross-like structures are observed in $\sigma^-$ polarization at a magnetic field $B_\mathrm{ac} =  3.75$ T (Fig. \ref{measurement}e,f). A similar spectral structure is visible for the biexciton (XX) at the same magnetic field, but in the opposite circular polarization $\sigma^+$. (Fig. \ref{measurement}a,b). The magnitudes of the splittings are identical due to their shared origin, explained by the cascade dynamics of the XX-X transition (Fig. \ref{levels}). The X levels involved in the anticrossing represent the final states after XX relaxation via the $\sigma^+$ optical transition, as well as the excited states before recombination through the $\sigma^-$ optical transition to a QD containing only the vanadium dopant, without additional carriers.

For the X$^{-}$, more anticrossings are observed, although the cross-like structure appears exclusively in $\sigma^-$ polarization (Fig. \ref{measurement}c,d). By combining these observations with the known g-factor signs of the V$^{2+}$ ion and the excitonic complexes, we determine that the hole-ion interaction is antiferromagnetic. This finding aligns with the behavior typical of II-VI diluted magnetic semiconductors \cite{Gaj2010}. Additionally, two anticrossings are detected at an energy of 1985.2,meV, visible in both $\sigma^+$ and $\sigma^-$ polarizations (Fig. \ref{measurement}c). Such features are characteristic of trions in solotronic systems \cite{Kudelski_2007_PRL, Smolenski_2017}, arising from the interaction of the ion with a single charge carrier in the QD.

However, in this case, the anticrossing observed in $\sigma^-$ is intertwined with one induced by shear strain effects on the valence band. This particular feature is not fully captured by the simplified model (Fig. \ref{measurement}d), suggesting the presence of a more intricate interaction between the vanadium dopant and the electron in the final state after X$^{-}$ recombination.

A detailed understanding of the interaction between vanadium and its crystal environment can be achieved by analyzing the energy levels of the neutral exciton, as illustrated in Fig. \ref{levels}. The excited state consists of an electron, a hole, and a vanadium ion within the quantum dot, while the ground state, following recombination, comprises only the vanadium ion in the quantum dot. The energy level analysis indicates that the observed anticrossings in the magnetic field are not solely attributable to the vanadium ion. Instead, they are significantly influenced by the hole in the excited state. Specifically, the component of the Hamiltonian associated with shear strain causes mixing of the hole states in the valence band. This mixing leads to the splitting of energy levels, corresponding to the same exciton configuration but different vanadium spin orientations in the magnetic field. The spin states ($\pm \frac{1}{2}$) of the vanadium ion do not interact directly, which accounts for the absence of anticrossings at zero magnetic field. However, under shear strain, these spin states become coupled to the excitonic complexes, enabling interactions that result in the observed anticrossings in the presence of a magnetic field.

Since no spectral lines corresponding to vanadium states with spin $\pm \frac{3}{2}$ are observed, nor any evidence of mixing between such states and the ground state, the analysis focuses exclusively on the ground state with spin $\pm \frac{1}{2}$. Higher-energy vanadium states with spin  $\pm \frac{3}{2}$ are not included. This represents a qualitatively distinct situation compared to a single Co$^{2+}$ ion in a QD \cite{kobak_2014_natcom,Kobak2018}, where the ground state is $\pm \frac{3}{2}$, and all spin states ($\pm \frac{3}{2}$ and $\pm \frac{1}{2}$) contribute to the excitonic spectrum, resulting in four lines at zero magnetic field even for a $\frac{1}{2}$ - $\frac{3}{2}$ splitting of approximately 3 meV.

Based on this, we conclude that in the case of vanadium in a CdTe QD, the $\frac{1}{2}$ - $\frac{3}{2}$ splitting must be significantly larger than that of cobalt. Such large splitting values are induced by the strain inherent to the QD. In bulk CdTe, the splitting between vanadium spin states ($\pm \frac{3}{2}$ and $\pm \frac{1}{2}$) is estimated to exceed 0.3 meV \cite{wanad_cdte_christmann}, whereas in bulk cobalt this splitting is negligible \cite{NAWROCKI1991111,PhysRevB.63.035202}.

To observe the higher-energy vanadium states in the future, two potential strategies could be employed. The first involves increasing the population of higher-energy states by raising the temperature or excitation power. However, both approaches also cause spectral line broadening, limiting their practical utility. The second strategy is to use very high magnetic fields. At sufficiently strong fields, the state with spin $\pm \frac{3}{2}$, which has a g-factor three times larger than that of $\pm \frac{1}{2}$, would become the ground state. This state could then be observed even at very low temperatures.

\section{Summary}

A novel solotronic system based on CdTe/ZnTe QDs doped with a single vanadium ion has been introduced. The shear strain present in these epitaxial nanostructures plays a significant role in modifying the valence band, giving rise to the key features observed in the optical spectra. Despite the relatively complex electronic and spin structure of V$^{2+}$ (spin $\frac{3}{2}$ with four possible spin projections), optical spectroscopy, supported by theoretical calculations, revealed only the degenerate  $\pm \frac{1}{2}$ spin states at zero magnetic field. Furthermore, the analysis confirmed that excited states of vanadium do not contribute to the ground state, even in the presence of an external magnetic field.

This simplicity, combined with the unique properties of these QDs, positions CdTe quantum dots containing a single vanadium ion as a promising candidate for qubit implementation in solid-state quantum information platforms.

\begin{acknowledgments}
The authors would like to thank Professor Lucienn Besombes for the discussion on the theoretical model. This work was partially supported by the Polish National Science Centre under Decisions No. DEC-2015/18/E/ST3/00559 and 2021/41/B/ST3/04183.
\end{acknowledgments}

\vskip\baselineskip

\bibliography{Bib_KP_QD_CdTe_V}

\begin{thebibliography}{41}
\expandafter\ifx\csname natexlab\endcsname\relax\def\natexlab#1{#1}\fi
\expandafter\ifx\csname bibnamefont\endcsname\relax
  \def\bibnamefont#1{#1}\fi
\expandafter\ifx\csname bibfnamefont\endcsname\relax
  \def\bibfnamefont#1{#1}\fi
\expandafter\ifx\csname citenamefont\endcsname\relax
  \def\citenamefont#1{#1}\fi
\expandafter\ifx\csname url\endcsname\relax
  \def\url#1{\texttt{#1}}\fi
\expandafter\ifx\csname urlprefix\endcsname\relax\def\urlprefix{URL }\fi
\providecommand{\bibinfo}[2]{#2}
\providecommand{\eprint}[2][]{\url{#2}}

\bibitem[{\citenamefont{Besombes et~al.}(2004)\citenamefont{Besombes,
  L{\'e}ger, Maingault, Ferrand, Mariette, and Cibert}}]{besombes_2004_prl}
\bibinfo{author}{\bibfnamefont{L.}~\bibnamefont{Besombes}},
  \bibinfo{author}{\bibfnamefont{Y.}~\bibnamefont{L{\'e}ger}},
  \bibinfo{author}{\bibfnamefont{L.}~\bibnamefont{Maingault}},
  \bibinfo{author}{\bibfnamefont{D.}~\bibnamefont{Ferrand}},
  \bibinfo{author}{\bibfnamefont{H.}~\bibnamefont{Mariette}}, \bibnamefont{and}
  \bibinfo{author}{\bibfnamefont{J.}~\bibnamefont{Cibert}},
  \emph{\bibinfo{title}{Probing the Spin State of a Single Magnetic Ion in an
  Individual Quantum Dot}}, \bibinfo{journal}{Physical Review Letters}
  \textbf{\bibinfo{volume}{93}}, \bibinfo{pages}{207403}
  (\bibinfo{year}{2004}),
  \urlprefix\url{https://doi.org/10.1103/PhysRevLett.93.207403}.

\bibitem[{\citenamefont{Kudelski et~al.}(2007)\citenamefont{Kudelski,
  Lema\^\i{}tre, Miard, Voisin, Graham, Warburton, and
  Krebs}}]{Kudelski_2007_PRL}
\bibinfo{author}{\bibfnamefont{A.}~\bibnamefont{Kudelski}},
  \bibinfo{author}{\bibfnamefont{A.}~\bibnamefont{Lema\^\i{}tre}},
  \bibinfo{author}{\bibfnamefont{A.}~\bibnamefont{Miard}},
  \bibinfo{author}{\bibfnamefont{P.}~\bibnamefont{Voisin}},
  \bibinfo{author}{\bibfnamefont{T.~C.~M.} \bibnamefont{Graham}},
  \bibinfo{author}{\bibfnamefont{R.~J.} \bibnamefont{Warburton}},
  \bibnamefont{and} \bibinfo{author}{\bibfnamefont{O.}~\bibnamefont{Krebs}},
  \emph{\bibinfo{title}{Optically Probing the Fine Structure of a Single Mn
  Atom in an InAs Quantum Dot}}, \bibinfo{journal}{Phys. Rev. Lett.}
  \textbf{\bibinfo{volume}{99}}, \bibinfo{pages}{247209}
  (\bibinfo{year}{2007}).

\bibitem[{\citenamefont{Le~Gall et~al.}(2009)\citenamefont{Le~Gall, Besombes,
  Boukari, Kolodka, Cibert, and Mariette}}]{LeGall_2009_PRL}
\bibinfo{author}{\bibfnamefont{C.}~\bibnamefont{Le~Gall}},
  \bibinfo{author}{\bibfnamefont{L.}~\bibnamefont{Besombes}},
  \bibinfo{author}{\bibfnamefont{H.}~\bibnamefont{Boukari}},
  \bibinfo{author}{\bibfnamefont{R.}~\bibnamefont{Kolodka}},
  \bibinfo{author}{\bibfnamefont{J.}~\bibnamefont{Cibert}}, \bibnamefont{and}
  \bibinfo{author}{\bibfnamefont{H.}~\bibnamefont{Mariette}},
  \emph{\bibinfo{title}{Optical Spin Orientation of a Single Manganese Atom in
  a Semiconductor Quantum Dot Using Quasiresonant Photoexcitation}},
  \bibinfo{journal}{Phys. Rev. Lett.} \textbf{\bibinfo{volume}{102}},
  \bibinfo{pages}{127402} (\bibinfo{year}{2009}).

\bibitem[{\citenamefont{Goryca et~al.}(2009)\citenamefont{Goryca, Kazimierczuk,
  Nawrocki, Golnik, Gaj, Kossacki, Wojnar, and Karczewski}}]{Goryca_2009_PRL2}
\bibinfo{author}{\bibfnamefont{M.}~\bibnamefont{Goryca}},
  \bibinfo{author}{\bibfnamefont{T.}~\bibnamefont{Kazimierczuk}},
  \bibinfo{author}{\bibfnamefont{M.}~\bibnamefont{Nawrocki}},
  \bibinfo{author}{\bibfnamefont{A.}~\bibnamefont{Golnik}},
  \bibinfo{author}{\bibfnamefont{J.~A.} \bibnamefont{Gaj}},
  \bibinfo{author}{\bibfnamefont{P.}~\bibnamefont{Kossacki}},
  \bibinfo{author}{\bibfnamefont{P.}~\bibnamefont{Wojnar}}, \bibnamefont{and}
  \bibinfo{author}{\bibfnamefont{G.}~\bibnamefont{Karczewski}},
  \emph{\bibinfo{title}{Optical Manipulation of a Single Mn Spin in a
  CdTe-Based Quantum Dot}}, \bibinfo{journal}{Phys. Rev. Lett.}
  \textbf{\bibinfo{volume}{103}}, \bibinfo{pages}{087401}
  (\bibinfo{year}{2009}).

\bibitem[{\citenamefont{Goryca et~al.}(2014)\citenamefont{Goryca, Koperski,
  Wojnar, Smole\'nski, Kazimierczuk, Golnik, and Kossacki}}]{Goryca_PRL_2014}
\bibinfo{author}{\bibfnamefont{M.}~\bibnamefont{Goryca}},
  \bibinfo{author}{\bibfnamefont{M.}~\bibnamefont{Koperski}},
  \bibinfo{author}{\bibfnamefont{P.}~\bibnamefont{Wojnar}},
  \bibinfo{author}{\bibfnamefont{T.}~\bibnamefont{Smole\'nski}},
  \bibinfo{author}{\bibfnamefont{T.}~\bibnamefont{Kazimierczuk}},
  \bibinfo{author}{\bibfnamefont{A.}~\bibnamefont{Golnik}}, \bibnamefont{and}
  \bibinfo{author}{\bibfnamefont{P.}~\bibnamefont{Kossacki}},
  \emph{\bibinfo{title}{Coherent precession of an individual 5/2 spin}},
  \bibinfo{journal}{Phys. Rev. Lett.} \textbf{\bibinfo{volume}{113}},
  \bibinfo{pages}{227202} (\bibinfo{year}{2014}).

\bibitem[{\citenamefont{Lafuente-Sampietro
  et~al.}(2015)\citenamefont{Lafuente-Sampietro, Boukari, and
  Besombes}}]{Lafuente_Sampietro_2015_PRB}
\bibinfo{author}{\bibfnamefont{A.}~\bibnamefont{Lafuente-Sampietro}},
  \bibinfo{author}{\bibfnamefont{H.}~\bibnamefont{Boukari}}, \bibnamefont{and}
  \bibinfo{author}{\bibfnamefont{L.}~\bibnamefont{Besombes}},
  \emph{\bibinfo{title}{Strain-induced coherent dynamics of coupled carriers
  and Mn spins in a quantum dot}}, \bibinfo{journal}{Phys. Rev. B}
  \textbf{\bibinfo{volume}{92}}, \bibinfo{pages}{081305}
  (\bibinfo{year}{2015}).

\bibitem[{\citenamefont{Fainblat et~al.}(2016)\citenamefont{Fainblat, Barrows,
  Hopmann, Siebeneicher, Vlaskin, Gamelin, and Bacher}}]{Fainblat_NL_2016}
\bibinfo{author}{\bibfnamefont{R.}~\bibnamefont{Fainblat}},
  \bibinfo{author}{\bibfnamefont{C.~J.} \bibnamefont{Barrows}},
  \bibinfo{author}{\bibfnamefont{E.}~\bibnamefont{Hopmann}},
  \bibinfo{author}{\bibfnamefont{S.}~\bibnamefont{Siebeneicher}},
  \bibinfo{author}{\bibfnamefont{V.~A.} \bibnamefont{Vlaskin}},
  \bibinfo{author}{\bibfnamefont{D.~R.} \bibnamefont{Gamelin}},
  \bibnamefont{and} \bibinfo{author}{\bibfnamefont{G.}~\bibnamefont{Bacher}},
  \emph{\bibinfo{title}{Giant Excitonic Exchange Splittings at Zero Field in
  Single Colloidal CdSe Quantum Dots Doped with Individual Mn2+ Impurities}},
  \bibinfo{journal}{Nano Lett.} \textbf{\bibinfo{volume}{16}},
  \bibinfo{pages}{6371} (\bibinfo{year}{2016}).

\bibitem[{\citenamefont{Kasprzak et~al.}(2022)\citenamefont{Kasprzak, Wigger,
  Hahn, Jakubczyk, Zinkiewicz, Machnikowski, Kuhn, Motte, and
  Pacuski}}]{Kasprzak2022}
\bibinfo{author}{\bibfnamefont{J.}~\bibnamefont{Kasprzak}},
  \bibinfo{author}{\bibfnamefont{D.}~\bibnamefont{Wigger}},
  \bibinfo{author}{\bibfnamefont{T.}~\bibnamefont{Hahn}},
  \bibinfo{author}{\bibfnamefont{T.}~\bibnamefont{Jakubczyk}},
  \bibinfo{author}{\bibfnamefont{L.}~\bibnamefont{Zinkiewicz}},
  \bibinfo{author}{\bibfnamefont{P.}~\bibnamefont{Machnikowski}},
  \bibinfo{author}{\bibfnamefont{T.}~\bibnamefont{Kuhn}},
  \bibinfo{author}{\bibfnamefont{J.-F.} \bibnamefont{Motte}}, \bibnamefont{and}
  \bibinfo{author}{\bibfnamefont{W.}~\bibnamefont{Pacuski}},
  \emph{\bibinfo{title}{Coherent Dynamics of a Single Mn-Doped Quantum Dot
  Revealed by Four-Wave Mixing Spectroscopy}}, \bibinfo{journal}{ACS Photonics}
  \textbf{\bibinfo{volume}{9}}, \bibinfo{pages}{1033} (\bibinfo{year}{2022}),
  \eprint{https://doi.org/10.1021/acsphotonics.1c01981},
  \urlprefix\url{https://doi.org/10.1021/acsphotonics.1c01981}.

\bibitem[{\citenamefont{Besombes et~al.}(2023)\citenamefont{Besombes, Ando,
  Kuroda, and Boukari}}]{PhysRevB.107.235305}
\bibinfo{author}{\bibfnamefont{L.}~\bibnamefont{Besombes}},
  \bibinfo{author}{\bibfnamefont{S.}~\bibnamefont{Ando}},
  \bibinfo{author}{\bibfnamefont{S.}~\bibnamefont{Kuroda}}, \bibnamefont{and}
  \bibinfo{author}{\bibfnamefont{H.}~\bibnamefont{Boukari}},
  \emph{\bibinfo{title}{Coupling of the triplet states of a negatively charged
  exciton in a quantum dot with the spin of a magnetic atom}},
  \bibinfo{journal}{Phys. Rev. B} \textbf{\bibinfo{volume}{107}},
  \bibinfo{pages}{235305} (\bibinfo{year}{2023}),
  \urlprefix\url{https://link.aps.org/doi/10.1103/PhysRevB.107.235305}.

\bibitem[{\citenamefont{Kobak et~al.}(2014)\citenamefont{Kobak, Smole{{\'
  n}}ski, Goryca, Papaj, Gietka, Bogucki, Koperski, Rousset, Suffczy{{\'
  n}}ski, Janik et~al.}}]{kobak_2014_natcom}
\bibinfo{author}{\bibfnamefont{J.}~\bibnamefont{Kobak}},
  \bibinfo{author}{\bibfnamefont{T.}~\bibnamefont{Smole{{\' n}}ski}},
  \bibinfo{author}{\bibfnamefont{M.}~\bibnamefont{Goryca}},
  \bibinfo{author}{\bibfnamefont{M.}~\bibnamefont{Papaj}},
  \bibinfo{author}{\bibfnamefont{K.}~\bibnamefont{Gietka}},
  \bibinfo{author}{\bibfnamefont{A.}~\bibnamefont{Bogucki}},
  \bibinfo{author}{\bibfnamefont{M.}~\bibnamefont{Koperski}},
  \bibinfo{author}{\bibfnamefont{J.-G.} \bibnamefont{Rousset}},
  \bibinfo{author}{\bibfnamefont{J.}~\bibnamefont{Suffczy{{\' n}}ski}},
  \bibinfo{author}{\bibfnamefont{E.}~\bibnamefont{Janik}},
  \bibinfo{author}{\bibfnamefont{M.}~\bibnamefont{Nawrocki}},
  \bibinfo{author}{\bibfnamefont{A.}~\bibnamefont{Golnik}},
  \bibinfo{author}{\bibfnamefont{P.}~\bibnamefont{Kossacki}}, \bibnamefont{and}
  \bibinfo{author}{\bibfnamefont{W.}~\bibnamefont{Pacuski}},
  \emph{\bibinfo{title}{Designing quantum dots for solotronics}},
  \bibinfo{journal}{Nature Communications} \textbf{\bibinfo{volume}{5}},
  \bibinfo{pages}{4191} (\bibinfo{year}{2014}),
  \urlprefix\url{https://doi.org/10.1038/ncomms4191}.

\bibitem[{\citenamefont{Kobak et~al.}(2018)\citenamefont{Kobak, Bogucki,
  Smole\ifmmode~\acute{n}\else \'{n}\fi{}ski, Papaj, Koperski, Potemski,
  Kossacki, Golnik, and Pacuski}}]{Kobak2018}
\bibinfo{author}{\bibfnamefont{J.}~\bibnamefont{Kobak}},
  \bibinfo{author}{\bibfnamefont{A.}~\bibnamefont{Bogucki}},
  \bibinfo{author}{\bibfnamefont{T.}~\bibnamefont{Smole\ifmmode~\acute{n}\else
  \'{n}\fi{}ski}}, \bibinfo{author}{\bibfnamefont{M.}~\bibnamefont{Papaj}},
  \bibinfo{author}{\bibfnamefont{M.}~\bibnamefont{Koperski}},
  \bibinfo{author}{\bibfnamefont{M.}~\bibnamefont{Potemski}},
  \bibinfo{author}{\bibfnamefont{P.}~\bibnamefont{Kossacki}},
  \bibinfo{author}{\bibfnamefont{A.}~\bibnamefont{Golnik}}, \bibnamefont{and}
  \bibinfo{author}{\bibfnamefont{W.}~\bibnamefont{Pacuski}},
  \emph{\bibinfo{title}{Direct determination of the zero-field splitting for a
  single ${\mathrm{Co}}^{2+}$ ion embedded in a CdTe/ZnTe quantum dot}},
  \bibinfo{journal}{Phys. Rev. B} \textbf{\bibinfo{volume}{97}},
  \bibinfo{pages}{045305} (\bibinfo{year}{2018}),
  \urlprefix\url{https://link.aps.org/doi/10.1103/PhysRevB.97.045305}.

\bibitem[{\citenamefont{Besombes et~al.}(2024)\citenamefont{Besombes, Kobak,
  and Pacuski}}]{PhysRevB.109.235302}
\bibinfo{author}{\bibfnamefont{L.}~\bibnamefont{Besombes}},
  \bibinfo{author}{\bibfnamefont{J.}~\bibnamefont{Kobak}}, \bibnamefont{and}
  \bibinfo{author}{\bibfnamefont{W.}~\bibnamefont{Pacuski}},
  \emph{\bibinfo{title}{Optical probing of the carrier-mediated coupling of the
  spin of two Co atoms in a quantum dot}}, \bibinfo{journal}{Phys. Rev. B}
  \textbf{\bibinfo{volume}{109}}, \bibinfo{pages}{235302}
  (\bibinfo{year}{2024}),
  \urlprefix\url{https://link.aps.org/doi/10.1103/PhysRevB.109.235302}.

\bibitem[{\citenamefont{Smole{{\' n}}ski et~al.}(2016)\citenamefont{Smole{{\'
  n}}ski, Kazimierczuk, Kobak, Goryca, Golnik, Kossacki, and
  Pacuski}}]{smolen_fe_2016_natcom}
\bibinfo{author}{\bibfnamefont{T.}~\bibnamefont{Smole{{\' n}}ski}},
  \bibinfo{author}{\bibfnamefont{T.}~\bibnamefont{Kazimierczuk}},
  \bibinfo{author}{\bibfnamefont{J.}~\bibnamefont{Kobak}},
  \bibinfo{author}{\bibfnamefont{M.}~\bibnamefont{Goryca}},
  \bibinfo{author}{\bibfnamefont{A.}~\bibnamefont{Golnik}},
  \bibinfo{author}{\bibfnamefont{P.}~\bibnamefont{Kossacki}}, \bibnamefont{and}
  \bibinfo{author}{\bibfnamefont{W.}~\bibnamefont{Pacuski}},
  \emph{\bibinfo{title}{Magnetic ground state of an individual Fe2+ ion in
  strained semiconductor nanostructure}}, \bibinfo{journal}{Nature
  Communications} \textbf{\bibinfo{volume}{7}}, \bibinfo{pages}{10484}
  (\bibinfo{year}{2016}), \urlprefix\url{https://doi.org/10.1038/ncomms10484}.

\bibitem[{\citenamefont{Smole\'{n}ski et~al.}(2017)\citenamefont{Smole\'{n}ski,
  Kazimierczuk, Goryca, Pacuski, and Kossacki}}]{Smolenski_2017}
\bibinfo{author}{\bibfnamefont{T.}~\bibnamefont{Smole\'{n}ski}},
  \bibinfo{author}{\bibfnamefont{T.}~\bibnamefont{Kazimierczuk}},
  \bibinfo{author}{\bibfnamefont{M.}~\bibnamefont{Goryca}},
  \bibinfo{author}{\bibfnamefont{W.}~\bibnamefont{Pacuski}}, \bibnamefont{and}
  \bibinfo{author}{\bibfnamefont{P.}~\bibnamefont{Kossacki}},
  \emph{\bibinfo{title}{Fine structure of an exciton coupled to a single
  Fe$^{2+}$ ion in a CdSe/ZnSe quantum dot}}, \bibinfo{journal}{Phys. Rev. B}
  \textbf{\bibinfo{volume}{96}}, \bibinfo{pages}{155411}
  (\bibinfo{year}{2017}).

\bibitem[{\citenamefont{Lafuente-Sampietro
  et~al.}(2016{\natexlab{a}})\citenamefont{Lafuente-Sampietro, Utsumi, Boukari,
  Kuroda, and Besombes}}]{Lafuente_2016}
\bibinfo{author}{\bibfnamefont{A.}~\bibnamefont{Lafuente-Sampietro}},
  \bibinfo{author}{\bibfnamefont{H.}~\bibnamefont{Utsumi}},
  \bibinfo{author}{\bibfnamefont{H.}~\bibnamefont{Boukari}},
  \bibinfo{author}{\bibfnamefont{S.}~\bibnamefont{Kuroda}}, \bibnamefont{and}
  \bibinfo{author}{\bibfnamefont{L.}~\bibnamefont{Besombes}},
  \emph{\bibinfo{title}{Individual Cr atom in a semiconductor quantum dot:
  Optical addressability and spin-strain coupling}}, \bibinfo{journal}{Phys.
  Rev. B} \textbf{\bibinfo{volume}{93}}, \bibinfo{pages}{161301}
  (\bibinfo{year}{2016}{\natexlab{a}}),
  \urlprefix\url{https://link.aps.org/doi/10.1103/PhysRevB.93.161301}.

\bibitem[{\citenamefont{Lafuente-Sampietro
  et~al.}(2016{\natexlab{b}})\citenamefont{Lafuente-Sampietro, Utsumi, Boukari,
  Kuroda, and Besombes}}]{Lafuente_APL_2016}
\bibinfo{author}{\bibfnamefont{A.}~\bibnamefont{Lafuente-Sampietro}},
  \bibinfo{author}{\bibfnamefont{H.}~\bibnamefont{Utsumi}},
  \bibinfo{author}{\bibfnamefont{H.}~\bibnamefont{Boukari}},
  \bibinfo{author}{\bibfnamefont{S.}~\bibnamefont{Kuroda}}, \bibnamefont{and}
  \bibinfo{author}{\bibfnamefont{L.}~\bibnamefont{Besombes}},
  \emph{\bibinfo{title}{Spin dynamics of an individual Cr atom in a
  semiconductor quantum dot under optical excitation}}, \bibinfo{journal}{Appl.
  Phys. Lett.} \textbf{\bibinfo{volume}{109}}, \bibinfo{eid}{053103}
  (\bibinfo{year}{2016}{\natexlab{b}}).

\bibitem[{\citenamefont{Besombes et~al.}(2019)\citenamefont{Besombes, Boukari,
  Tiwari, Lafuente-Sampietro, Sunaga, Makita, and Kuroda}}]{besombes_2019_prb}
\bibinfo{author}{\bibfnamefont{L.}~\bibnamefont{Besombes}},
  \bibinfo{author}{\bibfnamefont{H.}~\bibnamefont{Boukari}},
  \bibinfo{author}{\bibfnamefont{V.}~\bibnamefont{Tiwari}},
  \bibinfo{author}{\bibfnamefont{A.}~\bibnamefont{Lafuente-Sampietro}},
  \bibinfo{author}{\bibfnamefont{M.}~\bibnamefont{Sunaga}},
  \bibinfo{author}{\bibfnamefont{K.}~\bibnamefont{Makita}}, \bibnamefont{and}
  \bibinfo{author}{\bibfnamefont{S.}~\bibnamefont{Kuroda}},
  \emph{\bibinfo{title}{Charge fluctuations of a Cr atom probed in the optical
  spectra of a quantum dot}}, \bibinfo{journal}{Physical Review B}
  \textbf{\bibinfo{volume}{99}}, \bibinfo{pages}{035309}
  (\bibinfo{year}{2019}),
  \urlprefix\url{https://doi.org/10.1103/PhysRevB.99.035309}.

\bibitem[{\citenamefont{Tiwari et~al.}(2020)\citenamefont{Tiwari, Makita,
  Arino, Morita, Kuroda, Boukari, and Besombes}}]{Tiwari2020}
\bibinfo{author}{\bibfnamefont{V.}~\bibnamefont{Tiwari}},
  \bibinfo{author}{\bibfnamefont{K.}~\bibnamefont{Makita}},
  \bibinfo{author}{\bibfnamefont{M.}~\bibnamefont{Arino}},
  \bibinfo{author}{\bibfnamefont{M.}~\bibnamefont{Morita}},
  \bibinfo{author}{\bibfnamefont{S.}~\bibnamefont{Kuroda}},
  \bibinfo{author}{\bibfnamefont{H.}~\bibnamefont{Boukari}}, \bibnamefont{and}
  \bibinfo{author}{\bibfnamefont{L.}~\bibnamefont{Besombes}},
  \emph{\bibinfo{title}{Influence of nonequilibrium phonons on the spin
  dynamics of a single Cr atom}}, \bibinfo{journal}{Phys. Rev. B}
  \textbf{\bibinfo{volume}{101}}, \bibinfo{pages}{035305}
  (\bibinfo{year}{2020}),
  \urlprefix\url{https://link.aps.org/doi/10.1103/PhysRevB.101.035305}.

\bibitem[{\citenamefont{Tiwari et~al.}(2021{\natexlab{a}})\citenamefont{Tiwari,
  Arino, Gupta, Morita, Inoue, Caliste, Pochet, Boukari, Kuroda, and
  Besombes}}]{Tiwari2021}
\bibinfo{author}{\bibfnamefont{V.}~\bibnamefont{Tiwari}},
  \bibinfo{author}{\bibfnamefont{M.}~\bibnamefont{Arino}},
  \bibinfo{author}{\bibfnamefont{S.}~\bibnamefont{Gupta}},
  \bibinfo{author}{\bibfnamefont{M.}~\bibnamefont{Morita}},
  \bibinfo{author}{\bibfnamefont{T.}~\bibnamefont{Inoue}},
  \bibinfo{author}{\bibfnamefont{D.}~\bibnamefont{Caliste}},
  \bibinfo{author}{\bibfnamefont{P.}~\bibnamefont{Pochet}},
  \bibinfo{author}{\bibfnamefont{H.}~\bibnamefont{Boukari}},
  \bibinfo{author}{\bibfnamefont{S.}~\bibnamefont{Kuroda}}, \bibnamefont{and}
  \bibinfo{author}{\bibfnamefont{L.}~\bibnamefont{Besombes}},
  \emph{\bibinfo{title}{Hole-${\mathrm{Cr}}^{+}$ nanomagnet in a semiconductor
  quantum dot}}, \bibinfo{journal}{Phys. Rev. B}
  \textbf{\bibinfo{volume}{104}}, \bibinfo{pages}{L041301}
  (\bibinfo{year}{2021}{\natexlab{a}}),
  \urlprefix\url{https://link.aps.org/doi/10.1103/PhysRevB.104.L041301}.

\bibitem[{\citenamefont{Tiwari et~al.}(2022{\natexlab{a}})\citenamefont{Tiwari,
  Morita, Inoue, Ando, Kuroda, Boukari, and Besombes}}]{Tiwari2022}
\bibinfo{author}{\bibfnamefont{V.}~\bibnamefont{Tiwari}},
  \bibinfo{author}{\bibfnamefont{M.}~\bibnamefont{Morita}},
  \bibinfo{author}{\bibfnamefont{T.}~\bibnamefont{Inoue}},
  \bibinfo{author}{\bibfnamefont{S.}~\bibnamefont{Ando}},
  \bibinfo{author}{\bibfnamefont{S.}~\bibnamefont{Kuroda}},
  \bibinfo{author}{\bibfnamefont{H.}~\bibnamefont{Boukari}}, \bibnamefont{and}
  \bibinfo{author}{\bibfnamefont{L.}~\bibnamefont{Besombes}},
  \emph{\bibinfo{title}{Spin dynamics of positively charged excitons in
  ${\mathrm{Cr}}^{+}$-doped quantum dots probed by resonant
  photoluminescence}}, \bibinfo{journal}{Phys. Rev. B}
  \textbf{\bibinfo{volume}{106}}, \bibinfo{pages}{045308}
  (\bibinfo{year}{2022}{\natexlab{a}}),
  \urlprefix\url{https://link.aps.org/doi/10.1103/PhysRevB.106.045308}.

\bibitem[{\citenamefont{Tiwari et~al.}(2022{\natexlab{b}})\citenamefont{Tiwari,
  Morita, Inoue, Ando, Kuroda, Boukari, and Besombes}}]{PhysRevB.106.045308}
\bibinfo{author}{\bibfnamefont{V.}~\bibnamefont{Tiwari}},
  \bibinfo{author}{\bibfnamefont{M.}~\bibnamefont{Morita}},
  \bibinfo{author}{\bibfnamefont{T.}~\bibnamefont{Inoue}},
  \bibinfo{author}{\bibfnamefont{S.}~\bibnamefont{Ando}},
  \bibinfo{author}{\bibfnamefont{S.}~\bibnamefont{Kuroda}},
  \bibinfo{author}{\bibfnamefont{H.}~\bibnamefont{Boukari}}, \bibnamefont{and}
  \bibinfo{author}{\bibfnamefont{L.}~\bibnamefont{Besombes}},
  \emph{\bibinfo{title}{Spin dynamics of positively charged excitons in
  ${\mathrm{Cr}}^{+}$-doped quantum dots probed by resonant
  photoluminescence}}, \bibinfo{journal}{Phys. Rev. B}
  \textbf{\bibinfo{volume}{106}}, \bibinfo{pages}{045308}
  (\bibinfo{year}{2022}{\natexlab{b}}),
  \urlprefix\url{https://link.aps.org/doi/10.1103/PhysRevB.106.045308}.

\bibitem[{\citenamefont{Christmann
  et~al.}(1996{\natexlab{a}})\citenamefont{Christmann, Kreissl, Hofmann, Meyer,
  Schwarz, and Benz}}]{wanad_cdte_christmann}
\bibinfo{author}{\bibfnamefont{P.}~\bibnamefont{Christmann}},
  \bibinfo{author}{\bibfnamefont{J.}~\bibnamefont{Kreissl}},
  \bibinfo{author}{\bibfnamefont{D.}~\bibnamefont{Hofmann}},
  \bibinfo{author}{\bibfnamefont{B.}~\bibnamefont{Meyer}},
  \bibinfo{author}{\bibfnamefont{R.}~\bibnamefont{Schwarz}}, \bibnamefont{and}
  \bibinfo{author}{\bibfnamefont{K.}~\bibnamefont{Benz}},
  \emph{\bibinfo{title}{Vanadium in CdTe}}, \bibinfo{journal}{Journal of
  Crystal Growth} \textbf{\bibinfo{volume}{161}}, \bibinfo{pages}{259}
  (\bibinfo{year}{1996}{\natexlab{a}}),
  \urlprefix\url{https://www.sciencedirect.com/science/article/pii/0022024895006672}.

\bibitem[{\citenamefont{Peka et~al.}(1994)\citenamefont{Peka, Lehr, Schulz,
  Schwarz, and Benz}}]{wanad_cdte_peka}
\bibinfo{author}{\bibfnamefont{P.}~\bibnamefont{Peka}},
  \bibinfo{author}{\bibfnamefont{M.~U.} \bibnamefont{Lehr}},
  \bibinfo{author}{\bibfnamefont{H.~J.} \bibnamefont{Schulz}},
  \bibinfo{author}{\bibfnamefont{R.}~\bibnamefont{Schwarz}}, \bibnamefont{and}
  \bibinfo{author}{\bibfnamefont{K.~W.} \bibnamefont{Benz}},
  \emph{\bibinfo{title}{Energy levels of vanadium ions in CdTe}},
  \bibinfo{journal}{Applied Physics A} \textbf{\bibinfo{volume}{58}},
  \bibinfo{pages}{447–451} (\bibinfo{year}{1994}),
  \urlprefix\url{https://link.springer.com/article/10.1007/BF00332435}.

\bibitem[{\citenamefont{Christmann
  et~al.}(1996{\natexlab{b}})\citenamefont{Christmann, Meyer, Kreissl, Schwarz,
  and Benz}}]{PhysRevB.53.3634}
\bibinfo{author}{\bibfnamefont{P.}~\bibnamefont{Christmann}},
  \bibinfo{author}{\bibfnamefont{B.~K.} \bibnamefont{Meyer}},
  \bibinfo{author}{\bibfnamefont{J.}~\bibnamefont{Kreissl}},
  \bibinfo{author}{\bibfnamefont{R.}~\bibnamefont{Schwarz}}, \bibnamefont{and}
  \bibinfo{author}{\bibfnamefont{K.~W.} \bibnamefont{Benz}},
  \emph{\bibinfo{title}{Vanadium in CdTe: An electron-paramagnetic-resonance
  study}}, \bibinfo{journal}{Physical Review B} \textbf{\bibinfo{volume}{53}},
  \bibinfo{pages}{3634} (\bibinfo{year}{1996}{\natexlab{b}}),
  \urlprefix\url{https://link.aps.org/doi/10.1103/PhysRevB.53.3634}.

\bibitem[{\citenamefont{Schwartz et~al.}(1994)\citenamefont{Schwartz, Ziari,
  and Trivedi}}]{PhysRevB.49.5274}
\bibinfo{author}{\bibfnamefont{R.~N.} \bibnamefont{Schwartz}},
  \bibinfo{author}{\bibfnamefont{M.}~\bibnamefont{Ziari}}, \bibnamefont{and}
  \bibinfo{author}{\bibfnamefont{S.}~\bibnamefont{Trivedi}},
  \emph{\bibinfo{title}{Electron paramagnetic resonance and an optical
  investigation of photorefractive vanadium-doped CdTe}},
  \bibinfo{journal}{Physical Review B} \textbf{\bibinfo{volume}{49}},
  \bibinfo{pages}{5274} (\bibinfo{year}{1994}),
  \urlprefix\url{https://link.aps.org/doi/10.1103/PhysRevB.49.5274}.

\bibitem[{\citenamefont{Karczewski et~al.}(1999)\citenamefont{Karczewski,
  Ma\'{c}kowski, Kutrowski, Wojtowicz, and Kossut}}]{Karczewski1999}
\bibinfo{author}{\bibfnamefont{G.}~\bibnamefont{Karczewski}},
  \bibinfo{author}{\bibfnamefont{S.}~\bibnamefont{Ma\'{c}kowski}},
  \bibinfo{author}{\bibfnamefont{M.}~\bibnamefont{Kutrowski}},
  \bibinfo{author}{\bibfnamefont{T.}~\bibnamefont{Wojtowicz}},
  \bibnamefont{and} \bibinfo{author}{\bibfnamefont{J.}~\bibnamefont{Kossut}},
  \emph{\bibinfo{title}{{Photoluminescence study of CdTe/ZnTe self-assembled
  quantum dots}}}, \bibinfo{journal}{Applied Physics Letters}
  \textbf{\bibinfo{volume}{74}}, \bibinfo{pages}{3011} (\bibinfo{year}{1999}),
  ISSN \bibinfo{issn}{0003-6951},
  \eprint{https://pubs.aip.org/aip/apl/article-pdf/74/20/3011/7809986/3011\_1\_online.pdf},
  \urlprefix\url{https://doi.org/10.1063/1.123996}.

\bibitem[{\citenamefont{Ragunathan et~al.}(2019)\citenamefont{Ragunathan,
  Kobak, Gillard, Pacuski, Sobczak, Borysiuk, Skolnick, and
  Chekhovich}}]{Ragunathan2019}
\bibinfo{author}{\bibfnamefont{G.}~\bibnamefont{Ragunathan}},
  \bibinfo{author}{\bibfnamefont{J.}~\bibnamefont{Kobak}},
  \bibinfo{author}{\bibfnamefont{G.}~\bibnamefont{Gillard}},
  \bibinfo{author}{\bibfnamefont{W.}~\bibnamefont{Pacuski}},
  \bibinfo{author}{\bibfnamefont{K.}~\bibnamefont{Sobczak}},
  \bibinfo{author}{\bibfnamefont{J.}~\bibnamefont{Borysiuk}},
  \bibinfo{author}{\bibfnamefont{M.~S.} \bibnamefont{Skolnick}},
  \bibnamefont{and} \bibinfo{author}{\bibfnamefont{E.~A.}
  \bibnamefont{Chekhovich}}, \emph{\bibinfo{title}{Direct Measurement of
  Hyperfine Shifts and Radio Frequency Manipulation of Nuclear Spins in
  Individual $\mathrm{CdTe}/\mathrm{ZnTe}$ Quantum Dots}},
  \bibinfo{journal}{Phys. Rev. Lett.} \textbf{\bibinfo{volume}{122}},
  \bibinfo{pages}{096801} (\bibinfo{year}{2019}),
  \urlprefix\url{https://link.aps.org/doi/10.1103/PhysRevLett.122.096801}.

\bibitem[{\citenamefont{Tinjod et~al.}(2003)\citenamefont{Tinjod, Gilles,
  Moehl, Kheng, and Mariette}}]{Tinjod2003}
\bibinfo{author}{\bibfnamefont{F.}~\bibnamefont{Tinjod}},
  \bibinfo{author}{\bibfnamefont{B.}~\bibnamefont{Gilles}},
  \bibinfo{author}{\bibfnamefont{S.}~\bibnamefont{Moehl}},
  \bibinfo{author}{\bibfnamefont{K.}~\bibnamefont{Kheng}}, \bibnamefont{and}
  \bibinfo{author}{\bibfnamefont{H.}~\bibnamefont{Mariette}},
  \emph{\bibinfo{title}{{II–VI quantum dot formation induced by surface
  energy change of a strained layer}}}, \bibinfo{journal}{Applied Physics
  Letters} \textbf{\bibinfo{volume}{82}}, \bibinfo{pages}{4340}
  (\bibinfo{year}{2003}), ISSN \bibinfo{issn}{0003-6951},
  \eprint{https://pubs.aip.org/aip/apl/article-pdf/82/24/4340/12237803/4340\_1\_online.pdf},
  \urlprefix\url{https://doi.org/10.1063/1.1583141}.

\bibitem[{\citenamefont{Wojnar et~al.}(2007)\citenamefont{Wojnar, Karczewski,
  Wojtowicz, and Kossut}}]{Wojnar2007}
\bibinfo{author}{\bibfnamefont{P.}~\bibnamefont{Wojnar}},
  \bibinfo{author}{\bibfnamefont{G.}~\bibnamefont{Karczewski}},
  \bibinfo{author}{\bibfnamefont{T.}~\bibnamefont{Wojtowicz}},
  \bibnamefont{and} \bibinfo{author}{\bibfnamefont{J.}~\bibnamefont{Kossut}},
  \emph{\bibinfo{title}{Changing the Properties of the CdTe/ZnTe Quantum Dots
  by in situ Annealing during the Growth}}, \bibinfo{journal}{Acta Physica
  Polonica A} \textbf{\bibinfo{volume}{112}}, \bibinfo{pages}{283}
  (\bibinfo{year}{2007}).

\bibitem[{\citenamefont{Wojnar et~al.}(2011)\citenamefont{Wojnar, Bougerol,
  Bellet-Amalric, Besombes, Mariette, and Boukari}}]{wojnar-jcg-2011}
\bibinfo{author}{\bibfnamefont{P.}~\bibnamefont{Wojnar}},
  \bibinfo{author}{\bibfnamefont{C.}~\bibnamefont{Bougerol}},
  \bibinfo{author}{\bibfnamefont{E.}~\bibnamefont{Bellet-Amalric}},
  \bibinfo{author}{\bibfnamefont{L.}~\bibnamefont{Besombes}},
  \bibinfo{author}{\bibfnamefont{H.}~\bibnamefont{Mariette}}, \bibnamefont{and}
  \bibinfo{author}{\bibfnamefont{H.}~\bibnamefont{Boukari}},
  \emph{\bibinfo{title}{Towards vertical coupling of CdTe/ZnTe quantum dots
  formed by a high temperature tellurium induced process}},
  \bibinfo{journal}{J. Cryst. Growth} \textbf{\bibinfo{volume}{335}},
  \bibinfo{pages}{28} (\bibinfo{year}{2011}).

\bibitem[{\citenamefont{Kazimierczuk et~al.}(2013)\citenamefont{Kazimierczuk,
  Smole{{\' n}}ski, Kobak, Goryca, Pacuski, Golnik, Fronc, K{{\l}}opotowski,
  Wojnar, and Kossacki}}]{Kobak_statistics_2013}
\bibinfo{author}{\bibfnamefont{T.}~\bibnamefont{Kazimierczuk}},
  \bibinfo{author}{\bibfnamefont{T.}~\bibnamefont{Smole{{\' n}}ski}},
  \bibinfo{author}{\bibfnamefont{J.}~\bibnamefont{Kobak}},
  \bibinfo{author}{\bibfnamefont{M.}~\bibnamefont{Goryca}},
  \bibinfo{author}{\bibfnamefont{W.}~\bibnamefont{Pacuski}},
  \bibinfo{author}{\bibfnamefont{A.}~\bibnamefont{Golnik}},
  \bibinfo{author}{\bibfnamefont{K.}~\bibnamefont{Fronc}},
  \bibinfo{author}{\bibfnamefont{L.}~\bibnamefont{K{{\l}}opotowski}},
  \bibinfo{author}{\bibfnamefont{P.}~\bibnamefont{Wojnar}}, \bibnamefont{and}
  \bibinfo{author}{\bibfnamefont{P.}~\bibnamefont{Kossacki}},
  \emph{\bibinfo{title}{Optical study of electron-electron exchange interaction
  in CdTe/ZnTe quantum dots}}, \bibinfo{journal}{Physical Review B}
  \textbf{\bibinfo{volume}{87}}, \bibinfo{pages}{195302}
  (\bibinfo{year}{2013}),
  \urlprefix\url{https://journals.aps.org/prb/abstract/10.1103/PhysRevB.87.195302}.

\bibitem[{\citenamefont{Nawrocki et~al.}(1995)\citenamefont{Nawrocki, Rubo,
  Lascaray, and Coquillat}}]{Nawrocki1995}
\bibinfo{author}{\bibfnamefont{M.}~\bibnamefont{Nawrocki}},
  \bibinfo{author}{\bibfnamefont{Y.~G.} \bibnamefont{Rubo}},
  \bibinfo{author}{\bibfnamefont{J.~P.} \bibnamefont{Lascaray}},
  \bibnamefont{and}
  \bibinfo{author}{\bibfnamefont{D.}~\bibnamefont{Coquillat}},
  \emph{\bibinfo{title}{Suppression of the Auger recombination due to spin
  polarization of excess carriers and ${\mathrm{Mn}}^{2+}$ ions in the
  semimagnetic semiconductor ${\mathrm{Cd}}_{0.95}$${\mathrm{Mn}}_{0.05}$S}},
  \bibinfo{journal}{Phys. Rev. B} \textbf{\bibinfo{volume}{52}},
  \bibinfo{pages}{R2241} (\bibinfo{year}{1995}),
  \urlprefix\url{https://link.aps.org/doi/10.1103/PhysRevB.52.R2241}.

\bibitem[{\citenamefont{Jin et~al.}(2014)\citenamefont{Jin, Sun, Chen, Wei,
  Cheng, Li, and Li}}]{QuenchingVanadium}
\bibinfo{author}{\bibfnamefont{X.}~\bibnamefont{Jin}},
  \bibinfo{author}{\bibfnamefont{W.}~\bibnamefont{Sun}},
  \bibinfo{author}{\bibfnamefont{C.}~\bibnamefont{Chen}},
  \bibinfo{author}{\bibfnamefont{T.}~\bibnamefont{Wei}},
  \bibinfo{author}{\bibfnamefont{Y.}~\bibnamefont{Cheng}},
  \bibinfo{author}{\bibfnamefont{P.}~\bibnamefont{Li}}, \bibnamefont{and}
  \bibinfo{author}{\bibfnamefont{Q.}~\bibnamefont{Li}},
  \emph{\bibinfo{title}{Efficiency enhancement via tailoring energy level
  alignment induced by vanadium ion doping in organic/inorganic hybrid solar
  cells}}, \bibinfo{journal}{RSC Adv.} \textbf{\bibinfo{volume}{4}},
  \bibinfo{pages}{46008} (\bibinfo{year}{2014}),
  \urlprefix\url{http://dx.doi.org/10.1039/C4RA08671F}.

\bibitem[{\citenamefont{Oreszczuk et~al.}(2017)\citenamefont{Oreszczuk, Goryca,
  Pacuski, Smole\ifmmode~\acute{n}\else \'{n}\fi{}ski, Nawrocki, and
  Kossacki}}]{Oreszczuk2017}
\bibinfo{author}{\bibfnamefont{K.}~\bibnamefont{Oreszczuk}},
  \bibinfo{author}{\bibfnamefont{M.}~\bibnamefont{Goryca}},
  \bibinfo{author}{\bibfnamefont{W.}~\bibnamefont{Pacuski}},
  \bibinfo{author}{\bibfnamefont{T.}~\bibnamefont{Smole\ifmmode~\acute{n}\else
  \'{n}\fi{}ski}}, \bibinfo{author}{\bibfnamefont{M.}~\bibnamefont{Nawrocki}},
  \bibnamefont{and} \bibinfo{author}{\bibfnamefont{P.}~\bibnamefont{Kossacki}},
  \emph{\bibinfo{title}{Origin of luminescence quenching in structures
  containing CdSe/ZnSe quantum dots with a few ${\mathrm{Mn}}^{2+}$ ions}},
  \bibinfo{journal}{Phys. Rev. B} \textbf{\bibinfo{volume}{96}},
  \bibinfo{pages}{205436} (\bibinfo{year}{2017}),
  \urlprefix\url{https://link.aps.org/doi/10.1103/PhysRevB.96.205436}.

\bibitem[{\citenamefont{Piotrowski and Pacuski}(2020)}]{Piotrowski2020}
\bibinfo{author}{\bibfnamefont{P.}~\bibnamefont{Piotrowski}} \bibnamefont{and}
  \bibinfo{author}{\bibfnamefont{W.}~\bibnamefont{Pacuski}},
  \emph{\bibinfo{title}{Photoluminescence of CdTe quantum wells doped with
  cobalt and iron}}, \bibinfo{journal}{Journal of Luminescence}
  \textbf{\bibinfo{volume}{221}}, \bibinfo{pages}{117047}
  (\bibinfo{year}{2020}), ISSN \bibinfo{issn}{0022-2313},
  \urlprefix\url{https://www.sciencedirect.com/science/article/pii/S0022231319312785}.

\bibitem[{\citenamefont{Jasny et~al.}(1996)\citenamefont{Jasny, Sepiol,
  Irngartinger, Traber, Renn, and Wild}}]{jasny1996}
\bibinfo{author}{\bibfnamefont{J.}~\bibnamefont{Jasny}},
  \bibinfo{author}{\bibfnamefont{J.}~\bibnamefont{Sepiol}},
  \bibinfo{author}{\bibfnamefont{T.}~\bibnamefont{Irngartinger}},
  \bibinfo{author}{\bibfnamefont{M.}~\bibnamefont{Traber}},
  \bibinfo{author}{\bibfnamefont{A.}~\bibnamefont{Renn}}, \bibnamefont{and}
  \bibinfo{author}{\bibfnamefont{U.~P.} \bibnamefont{Wild}},
  \emph{\bibinfo{title}{Fluorescence microscopy in superfluid helium: Single
  molecule imaging}}, \bibinfo{journal}{Review of Scientific Instruments}
  \textbf{\bibinfo{volume}{67}} (\bibinfo{year}{1996}),
  \urlprefix\url{https://doi.org/10.1063/1.1146868}.

\bibitem[{\citenamefont{Goryca et~al.}(2010)\citenamefont{Goryca,
  P{{\l}}ochocka, Kazimierczuk, Wojnar, Karczewski, Gaj, Potemski, and
  Kossacki}}]{Goryca_PRB_dark}
\bibinfo{author}{\bibfnamefont{M.}~\bibnamefont{Goryca}},
  \bibinfo{author}{\bibfnamefont{P.}~\bibnamefont{P{{\l}}ochocka}},
  \bibinfo{author}{\bibfnamefont{T.}~\bibnamefont{Kazimierczuk}},
  \bibinfo{author}{\bibfnamefont{P.}~\bibnamefont{Wojnar}},
  \bibinfo{author}{\bibfnamefont{G.}~\bibnamefont{Karczewski}},
  \bibinfo{author}{\bibfnamefont{J.~A.} \bibnamefont{Gaj}},
  \bibinfo{author}{\bibfnamefont{M.}~\bibnamefont{Potemski}}, \bibnamefont{and}
  \bibinfo{author}{\bibfnamefont{P.}~\bibnamefont{Kossacki}},
  \emph{\bibinfo{title}{Brightening of dark excitons in a single CdTe quantum
  dot containing a single ${\text{Mn}}^{2+}$ ion}}, \bibinfo{journal}{Phys.
  Rev. B} \textbf{\bibinfo{volume}{82}}, \bibinfo{pages}{165323}
  (\bibinfo{year}{2010}),
  \urlprefix\url{https://link.aps.org/doi/10.1103/PhysRevB.82.165323}.

\bibitem[{\citenamefont{Tiwari et~al.}(2021{\natexlab{b}})\citenamefont{Tiwari,
  Arino, Gupta, Morita, Inoue, Caliste, Pochet, Boukari, Kuroda, and
  Besombes}}]{PhysRevB.104.L041301}
\bibinfo{author}{\bibfnamefont{V.}~\bibnamefont{Tiwari}},
  \bibinfo{author}{\bibfnamefont{M.}~\bibnamefont{Arino}},
  \bibinfo{author}{\bibfnamefont{S.}~\bibnamefont{Gupta}},
  \bibinfo{author}{\bibfnamefont{M.}~\bibnamefont{Morita}},
  \bibinfo{author}{\bibfnamefont{T.}~\bibnamefont{Inoue}},
  \bibinfo{author}{\bibfnamefont{D.}~\bibnamefont{Caliste}},
  \bibinfo{author}{\bibfnamefont{P.}~\bibnamefont{Pochet}},
  \bibinfo{author}{\bibfnamefont{H.}~\bibnamefont{Boukari}},
  \bibinfo{author}{\bibfnamefont{S.}~\bibnamefont{Kuroda}}, \bibnamefont{and}
  \bibinfo{author}{\bibfnamefont{L.}~\bibnamefont{Besombes}},
  \emph{\bibinfo{title}{Hole-${\mathrm{Cr}}^{+}$ nanomagnet in a semiconductor
  quantum dot}}, \bibinfo{journal}{Physical Review B}
  \textbf{\bibinfo{volume}{104}}, \bibinfo{pages}{L041301}
  (\bibinfo{year}{2021}{\natexlab{b}}),
  \urlprefix\url{https://link.aps.org/doi/10.1103/PhysRevB.104.L041301}.

\bibitem[{\citenamefont{Gaj and Kossut}(2010)}]{Gaj2010}
\bibinfo{author}{\bibfnamefont{J.~A.} \bibnamefont{Gaj}} \bibnamefont{and}
  \bibinfo{author}{\bibfnamefont{J.}~\bibnamefont{Kossut}},
  \emph{\bibinfo{title}{Introduction to the physics of diluted magnetic
  semiconductors}}, vol. \bibinfo{volume}{144} (\bibinfo{publisher}{Springer
  Science \& Business Media}, \bibinfo{year}{2010}).

\bibitem[{\citenamefont{Nawrocki et~al.}(1991)\citenamefont{Nawrocki, Hamdani,
  Lascaray, Golacki, and Deportes}}]{NAWROCKI1991111}
\bibinfo{author}{\bibfnamefont{M.}~\bibnamefont{Nawrocki}},
  \bibinfo{author}{\bibfnamefont{F.}~\bibnamefont{Hamdani}},
  \bibinfo{author}{\bibfnamefont{J.}~\bibnamefont{Lascaray}},
  \bibinfo{author}{\bibfnamefont{Z.}~\bibnamefont{Golacki}}, \bibnamefont{and}
  \bibinfo{author}{\bibfnamefont{J.}~\bibnamefont{Deportes}},
  \emph{\bibinfo{title}{Ion-carrier electron exchange constants for CdCoSe
  semimagnetic semiconductor}}, \bibinfo{journal}{Solid State Communications}
  \textbf{\bibinfo{volume}{77}}, \bibinfo{pages}{111} (\bibinfo{year}{1991}),
  ISSN \bibinfo{issn}{0038-1098},
  \urlprefix\url{https://www.sciencedirect.com/science/article/pii/003810989190868V}.

\bibitem[{\citenamefont{Zieli{{\' n}}ski et~al.}(2000)\citenamefont{Zieli{{\'
  n}}ski, Rigaux, Mycielski, and Menant}}]{PhysRevB.63.035202}
\bibinfo{author}{\bibfnamefont{M.}~\bibnamefont{Zieli{{\' n}}ski}},
  \bibinfo{author}{\bibfnamefont{C.}~\bibnamefont{Rigaux}},
  \bibinfo{author}{\bibfnamefont{A.}~\bibnamefont{Mycielski}},
  \bibnamefont{and} \bibinfo{author}{\bibfnamefont{M.}~\bibnamefont{Menant}},
  \emph{\bibinfo{title}{Zeeman spectrum of the $1s$ exciton in very diluted
  ${\mathrm{Cd}}_{1\ensuremath{-}x}{\mathrm{Co}}_{x}\mathrm{Te}$ compounds}},
  \bibinfo{journal}{Phys. Rev. B} \textbf{\bibinfo{volume}{63}},
  \bibinfo{pages}{035202} (\bibinfo{year}{2000}),
  \urlprefix\url{https://link.aps.org/doi/10.1103/PhysRevB.63.035202}.

\end{thebibliography}


\begin{thebibliography}{1}
\expandafter\ifx\csname natexlab\endcsname\relax\def\natexlab#1{#1}\fi
\expandafter\ifx\csname bibnamefont\endcsname\relax
  \def\bibnamefont#1{#1}\fi
\expandafter\ifx\csname bibfnamefont\endcsname\relax
  \def\bibfnamefont#1{#1}\fi
\expandafter\ifx\csname citenamefont\endcsname\relax
  \def\citenamefont#1{#1}\fi
\expandafter\ifx\csname url\endcsname\relax
  \def\url#1{\texttt{#1}}\fi
\expandafter\ifx\csname urlprefix\endcsname\relax\def\urlprefix{URL }\fi
\providecommand{\bibinfo}[2]{#2}
\providecommand{\eprint}[2][]{\url{#2}}

\bibitem[{\citenamefont{Tiwari et~al.}(2021)\citenamefont{Tiwari, Arino, Gupta,
  Morita, Inoue, Caliste, Pochet, Boukari, Kuroda, and Besombes}}]{Tiwari2021}
\bibinfo{author}{\bibfnamefont{V.}~\bibnamefont{Tiwari}},
  \bibinfo{author}{\bibfnamefont{M.}~\bibnamefont{Arino}},
  \bibinfo{author}{\bibfnamefont{S.}~\bibnamefont{Gupta}},
  \bibinfo{author}{\bibfnamefont{M.}~\bibnamefont{Morita}},
  \bibinfo{author}{\bibfnamefont{T.}~\bibnamefont{Inoue}},
  \bibinfo{author}{\bibfnamefont{D.}~\bibnamefont{Caliste}},
  \bibinfo{author}{\bibfnamefont{P.}~\bibnamefont{Pochet}},
  \bibinfo{author}{\bibfnamefont{H.}~\bibnamefont{Boukari}},
  \bibinfo{author}{\bibfnamefont{S.}~\bibnamefont{Kuroda}}, \bibnamefont{and}
  \bibinfo{author}{\bibfnamefont{L.}~\bibnamefont{Besombes}},
  \emph{\bibinfo{title}{Hole-${\mathrm{Cr}}^{+}$ nanomagnet in a semiconductor
  quantum dot}}, \bibinfo{journal}{Phys. Rev. B}
  \textbf{\bibinfo{volume}{104}}, \bibinfo{pages}{L041301}
  (\bibinfo{year}{2021}),
  \urlprefix\url{https://link.aps.org/doi/10.1103/PhysRevB.104.L041301}.

\end{thebibliography}

\bibliographystyle{apsrev_my}

\end{document}


\title{Single vanadium ion magnetic dopant in an individual CdTe/ZnTe quantum dot \\ Supplementary Materials}

\author{K. E. \surname{Po\l{}czy\'nska}}\affiliation{\FUW}\email{kpolczynska@fuw.edu.pl, k.e.polczynska@gmail.com}
\author{T. \surname{Kazimierczuk}}\affiliation{\FUW}
\author{P. \surname{Kossacki}}\affiliation{\FUW}
\author{W. \surname{Pacuski}}\affiliation{\FUW}

\date{\today}

\maketitle

\section{Theoretical model}

Hamiltonian of the neutral exciton interactin with vanadium ion inside the
QD was written using the base of $ (| \Uparrow \downarrow >, |\Downarrow
\uparrow>)\otimes(| \frac{1}{2} >, | - \frac{1}{2} > )  = |  \Uparrow
\downarrow ,\frac{1}{2} >, |  \Downarrow \uparrow, \frac{1}{2} >, |
\Uparrow \downarrow, -\frac{1}{2} >, | \Downarrow \uparrow, -\frac{1}{2}
>$, where $| \frac{1}{2} >, | -\frac{1}{2} >$ represented the spin of an
ion and $| \Uparrow \downarrow >, | \Downarrow \uparrow >$ pictured the
spin of the exciton.

Before the recombination, such system can be described by:

\begin{equation}
 \begin{multlined}
H_{X+V}=\frac{1}{2}
\begin{pmatrix}
\Delta_{s,p-d} & \delta_X & 0 & 0
\\
\delta_X & -\Delta_{s,p-d} & 0 &
0\\
0 & 0 & -\Delta_{s,p-d}& \delta_X
\\
0 & 0 & \delta_X & \Delta_{s,p-d }
\end{pmatrix}_{1,2}
+\frac{1}{2}
\begin{pmatrix}
\ - \mu_bg_XB  & 0 & 0 & 0
\\
0 & \mu_bg_XB  & 0 & 0
\\
0 & 0 & - \mu_bg_XB  & 0
\\
0 & 0 & 0 & \mu_bg_XB 
\end{pmatrix}_3 +
\\+\frac{1}{2}
\begin{pmatrix}
\  - \mu_bg_{V}B & 0 & 0 & 0
\\
0 &  - \mu_bg_{V}B & 0 & 0
\\
0 & 0 &  \mu_bg_{V}B & 0
\\
0 & 0 & 0 &  \mu_bg_{V}B
\end{pmatrix}_4
+\frac{1}{2}
\begin{pmatrix}
\  \gamma B^2 & 0 & 0 & 0
\\
0 & \gamma B^2 & 0 & 0
\\
 0 & 0 &  \gamma B^2 & 0
\\
0 & 0 & 0 &  \gamma B^2
\end{pmatrix}_5
+ \frac{2}{3}\Delta_{p-d} 
\begin{pmatrix}
\  0 & 0 & \sqrt{3}\xi & 0
\\
0 & 0 & 0 & - \sqrt{3}\xi
\\
 \sqrt{3}\xi^* & 0 &  0 & 0
\\
0 & - \sqrt{3}\xi^* & 0 &  0
\end{pmatrix}_6  
 \end{multlined}
\label{eq:hamX+ion_2}
\end{equation}

and after recombination, the system consisting of single vanadium and QD
without any carriers can be described as:

\begin{equation}
H_{V} = \frac{1}{2}
\begin{pmatrix}
\ -\mu_bg_{V}B & 0 
\\
\ 0 &  \mu_bg_{V}B  
\end{pmatrix}
\label{eq:ham_ion_2}
\end{equation}

where $\Delta_{s,p-d}$ and $\Delta_{p-d}$ are constants
describing exciton-ion and hole-ion exchange interactions,
$\delta_X$ and $\delta_V$ are constants representing the splitting of
energy levels due to the anisotropy of the QD, $\gamma$ is an excitonic
diamagnetic shift constant, $\mu_B$ is a Bohr magneton, $g_X$ and $g_V$ are
g-factors of the exciton and the vanadium and $\xi$ is the parameter
describing the valence band mixing due to the sheer strain. Spin operators
of the ion ($S$) and exciton ($\sigma$) are taken into account in their
usual form.

Part $H_{6}$ of the Hamiltonian is describing the influence of the shear
strain on the QD with V$^{2+}$ dopant and was calculated the following
way:

\begin{equation}
H_{6} = \frac{2}{3}\Delta_{p-d} (S_+ \otimes \sigma_- + S_- \otimes
\sigma_+ )
\label{poprawka_wzor}
\end{equation}

for spin $\frac{1}{2}$ of vanadium:

\begin{equation*}
S_z = 
\begin{pmatrix}
\frac{1}{2} & 0
\\
0 & - \frac{1}{2}
\end{pmatrix},
S_+ = 
\begin{pmatrix}
0 & 0
\\
1 & 0
\end{pmatrix},
S_- = 
\begin{pmatrix}
0 & 1
\\
0 & 0
\end{pmatrix}.
\end{equation*}

According to supplementary information of \cite{Tiwari2021}, $J$
operators in presence of shear strain have form of:

\begin{equation*}
\sigma_z = 
\begin{pmatrix}
\frac{3}{2} & 0
\\
0 & -\frac{3}{2}
\end{pmatrix},
\sigma_+ = \xi
\begin{pmatrix}
\sqrt{3} & 0
\\
0 & -\sqrt{3}
\end{pmatrix},
\sigma_- = \xi^{*}
\begin{pmatrix}
\sqrt{3} & 0
\\
0 & -\sqrt{3}
\end{pmatrix},
\end{equation*}

so after calculating the equation \ref{poprawka_wzor} we get the final form
of the $H_{6}$ as:

\begin{equation}
H_{6} = \frac{2}{3}\Delta_{p-d} 
\begin{pmatrix}
\  0 & 0 & \sqrt{3}\xi & 0
\\
0 & 0 & 0 & - \sqrt{3}\xi
\\
 \sqrt{3}\xi^* & 0 &  0 & 0
\\
0 & - \sqrt{3}\xi^* & 0 &  0
\end{pmatrix}.
\end{equation}

\section{Experiment}

\begin{figure*}[h!]
 \renewcommand{\thefigure}{S1}
\includegraphics[width=0.75\textwidth]{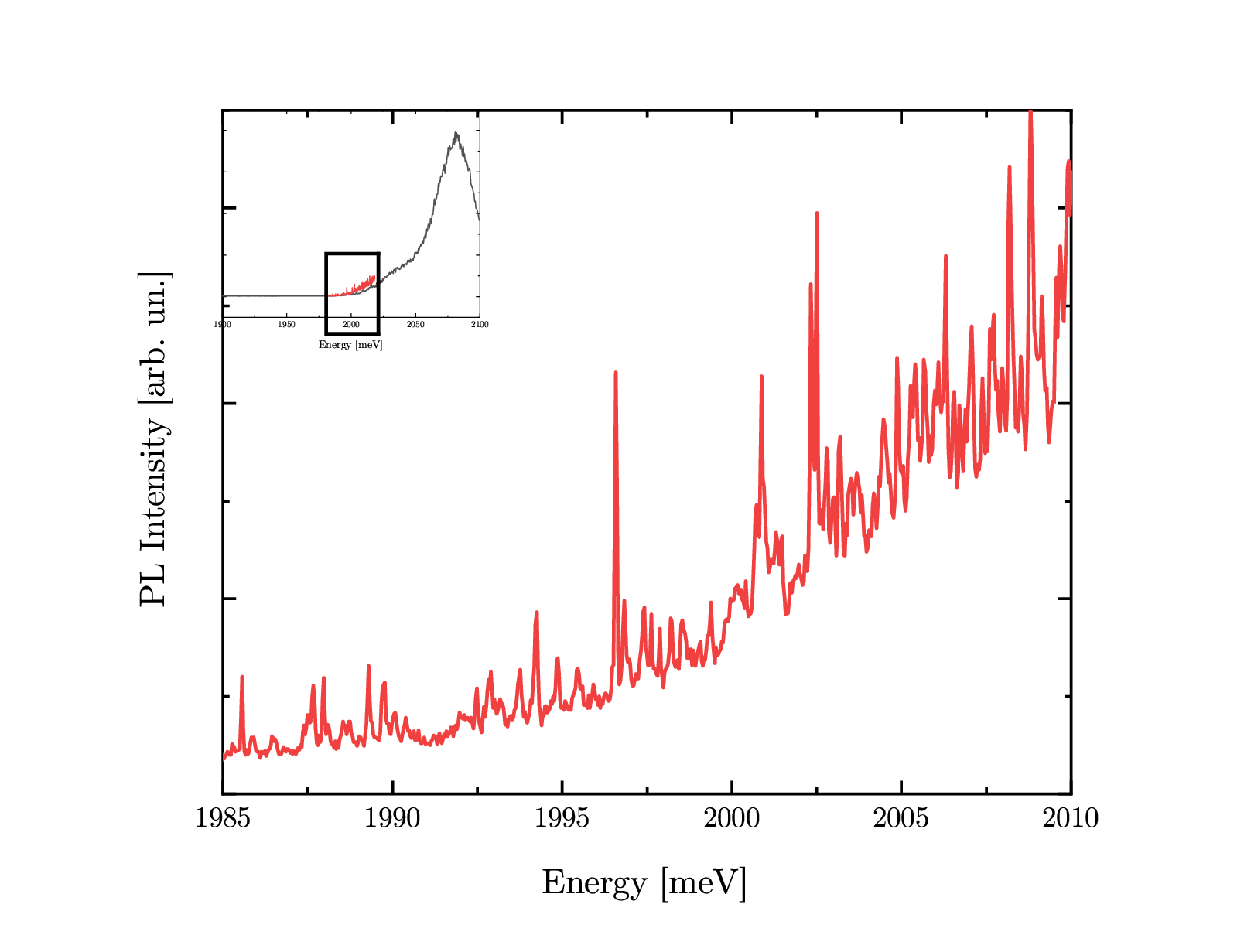}
\caption{\textbf{Microphotoluminescence measured at helium temperature of 10 K.} In the low energetic part of the spectrum one can observe single emission lines originating from single quantum dots.}
\label{patelnia}
\end{figure*}

\begin{figure*}
 \renewcommand{\thefigure}{S2}
\includegraphics[width=\textwidth]{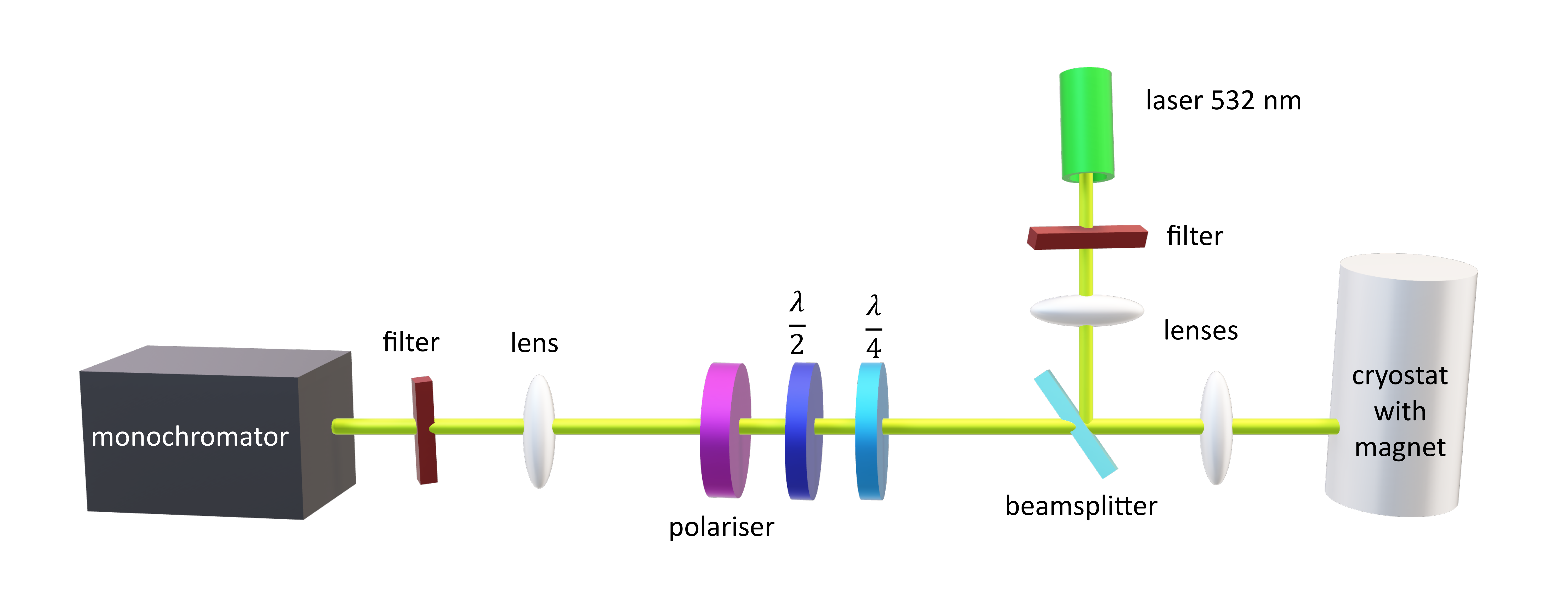}
\caption{\textbf{Experimental setup dedicated to polarization-resolved magnetospectroscopy measurement.} Sample was excited with 532 nm LED laser. The same setup, but without the $\frac{\lambda}{4}$ waveplate was used to measure the angular dependence of the QD spectrum.}
\label{setup}
\end{figure*}

\begin{figure*}
 \renewcommand{\thefigure}{S3}
\includegraphics[width=\textwidth]{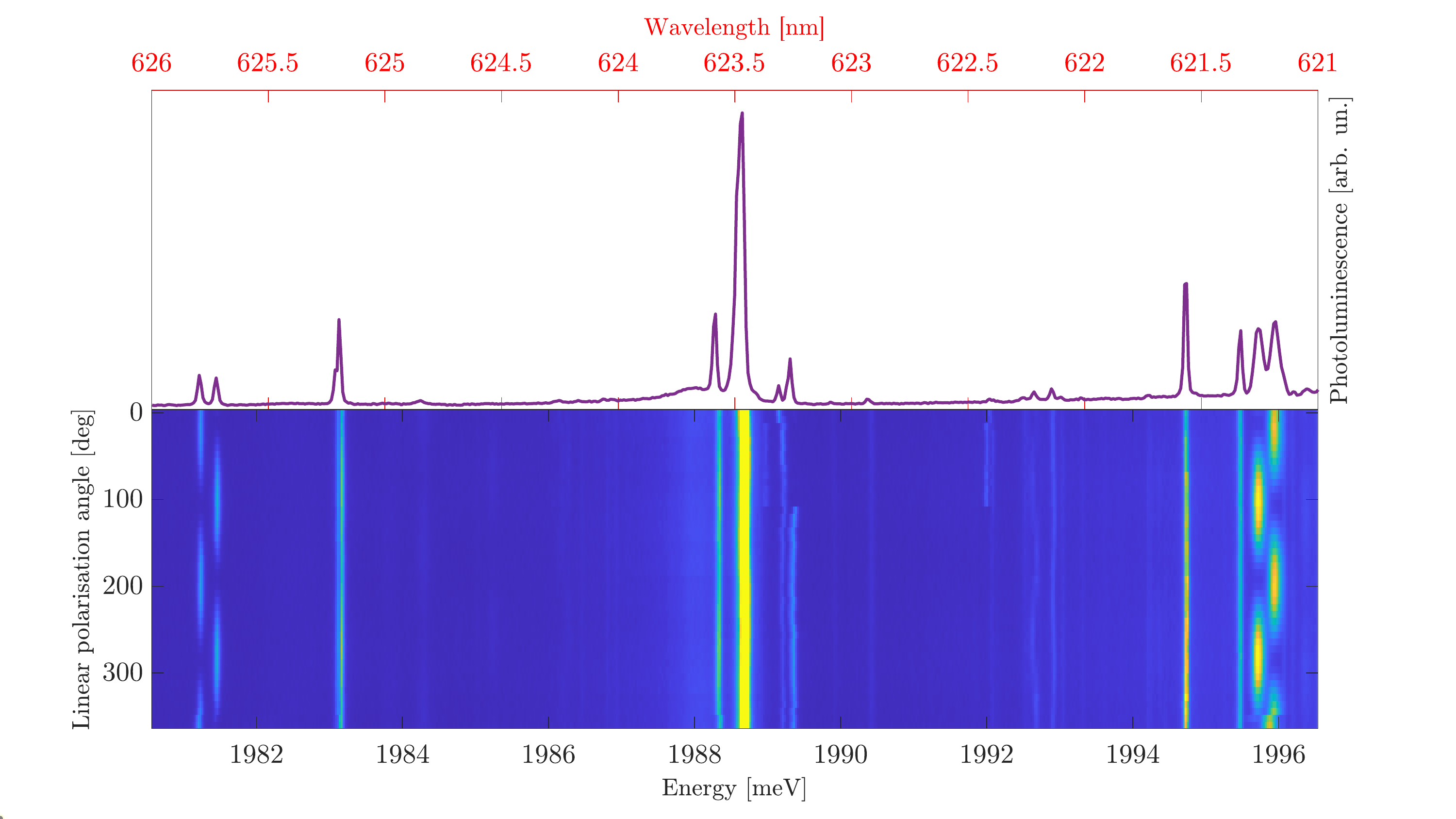}
\caption{\textbf{Anisotropy of the regular, undoped CdTe/ZnTe QD.} Linear polarization angular dependence of photoluminescence measured at 5 K. X and XX are fully orthogonally polarized to each other, while trions do not exhibit any intensity dependence on the linear polarization of detection.}
\label{anizo_casual}
\end{figure*}

\begin{figure*}
 \renewcommand{\thefigure}{S4}
\includegraphics[width=\textwidth]{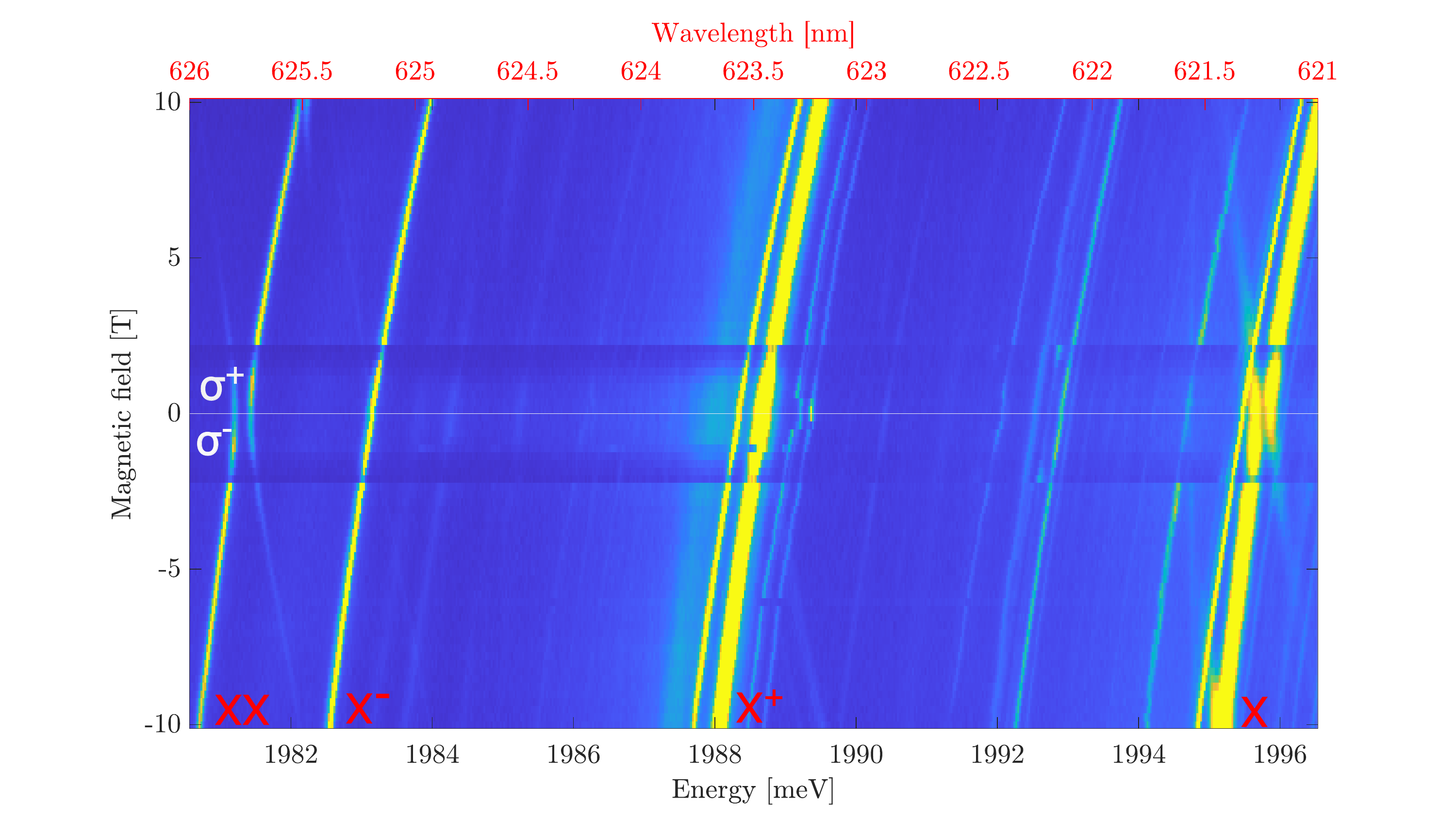}
\caption{\textbf{Magnetic dependence of the photoluminescence spectrum of the regular, undoped CdTe/ZnTe QD.} Anticrossing on X and XX in B = 0 T is visible, when their two components separate for two different circular polarization with the application of the magnetic field. No anticrossings were observed for trions. Excitonic lines are affected mainly by Zeemann effect.}
\label{setup}
\end{figure*}

\begin{figure*}
 \renewcommand{\thefigure}{S5}
\includegraphics[width=\textwidth]{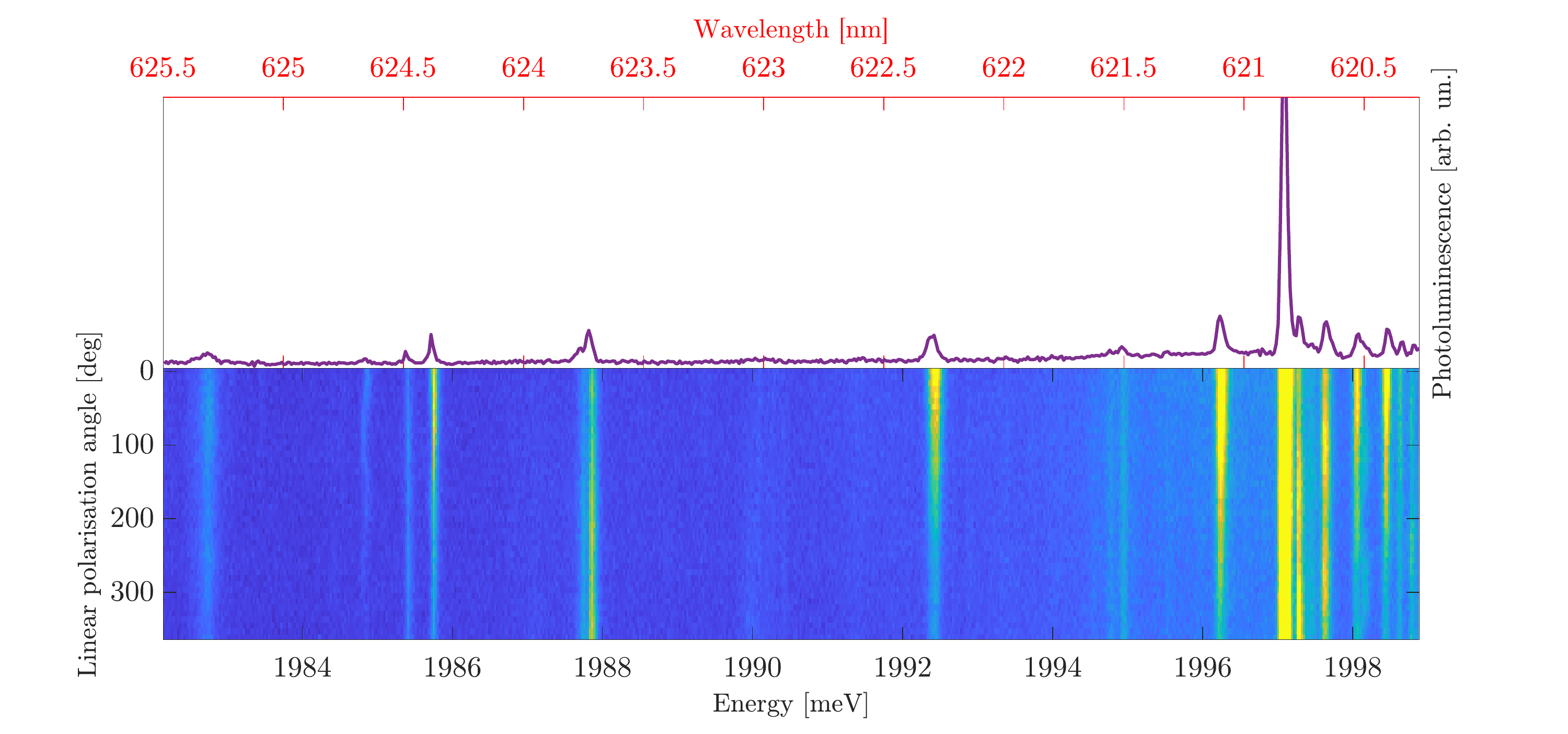}
\caption{\textbf{Anisotropy of the CdTe/ZnTe QD with a single Vanadium dopant.} Linear polarization angular dependence of photoluminescence measured at 5 K. No influence on the intensity of particular peaks has been observed.}
\label{setup}
\end{figure*}
	

\bibliography{Bib_KP_QD_CdTe_V}

\bibliographystyle{apsrev_my}